\documentclass[
 reprint,
 amsmath,amssymb,
 aps,superscriptaddress,prl
]{revtex4-2}

\usepackage{dcolumn}
\usepackage{bm}
\usepackage{physics}
\usepackage{caption}
\usepackage{amsfonts}
\usepackage[utf8]{inputenc}
\usepackage{chemformula}
\usepackage{amsmath}
\usepackage{dsfont}
\usepackage{float}
\usepackage{enumitem}
\usepackage[bb=boondox]{mathalfa}
\newcommand{\kB}{k_\mathrm{B}}
\newcommand{\ltf}{\ell_\mathrm{TF}}

\begin{document}
\preprint{1}

\title{Microscopic origin of twist-dependent electron transfer rate in bilayer graphene}
\author{Leonardo Coello Escalante}
\affiliation{%
 Department of Chemistry, University of California, Berkeley, CA 94720, USA
}%

\author{David T. Limmer}
 \email{dlimmer@berkeley.edu}
\affiliation{%
 Department of Chemistry, University of California, Berkeley, CA 94720, USA
}%
\affiliation{%
 Kavli Energy NanoScience Institute, Berkeley, CA 94720, USA
}%
\affiliation{%
 MSD, Lawrence Berkeley National Laboratory, Berkeley, CA 94720, USA
}%
\affiliation{%
 CSD, Lawrence Berkeley, National Laboratory, Berkeley, CA 94720, USA
}%

\date{\today}

\begin{abstract}
\textbf{Abstract:} Using molecular simulation and continuum dielectric theory, we consider how electrochemical kinetics are modulated by twist angle in bilayer graphene electrodes. By establishing a connection between twist angle and the screening length of charge carriers within the electrode, we investigate how tunable metallicity modifies the statistics of the electron transfer energy gap. Constant potential molecular simulations show that the activation free energy for electron transfer increases with screening length, leading to a non-monotonic dependence on the twist angle. The twist angle alters the density of states, tuning the number of thermally-accessible channels for electron transfer and the reorganization energy by affecting the stability of the vertically excited state through attenuated image charge interactions. Understanding these effects allows us to express the Marcus rate of interfacial electron transfer as a function of twist angle, consistent with a growing body of experimental observations. \textbf{Keywords:} Electrochemistry, Twisted Bilayer Graphene, Electron Transfer, Marcus Theory, Two-Dimensional Materials
\end{abstract}

\maketitle


Two-dimensional materials are increasingly important platforms for probing fundamental processes in condensed matter, as well as for the design of novel technologies \cite{fiori_electronics_2014, jin_emerging_2018, chia_characteristics_2018}. Van der Waals materials in particular offer the opportunity to modify electronic interactions with unprecedented freedom by changing the identity of the layers, mechanically modifying their stacking configuration, or electrostatically gating them \cite{geim_van_2013}.  Moiré superlattices can be achieved in such materials by introducing a relative twist angle between bilayers\cite{andrei_marvels_2021, he_moire_2021}, and have been extensively studied in twisted bilayer graphene (TBG) \cite{nimbalkar_opportunities_2020, andrei_graphene_2020}. Special twist angles known as `magic angles', result in the formation of essentially flat electronic bands near the Fermi level, and can contribute to the emergence of exotic electronic phenomena at low temperature\cite{cao_unconventional_2018, cao_correlated_2018, yankowitz_tuning_2019, choi_imaging_2019}. Efficient electrochemical interfaces often require tailoring the electronic structure of the electrode to enhance electron transfer over the course of thermal fluctuations \cite{seh_combining_2017, stamenkovic_energy_2017, sundararaman2022improving}, so the electronic tunability of TBG presents the tantalizing opportunity to control electron transfer kinetics\cite{deng_catalysis_2016, li_moire_2023}. Indeed, twist angle-dependent rates of interfacial electron transfer have been experimentally observed in TBG electrodes \cite{yu_tunable_2022}. 

\begin{figure}[t]
    \includegraphics[width=8.25cm]{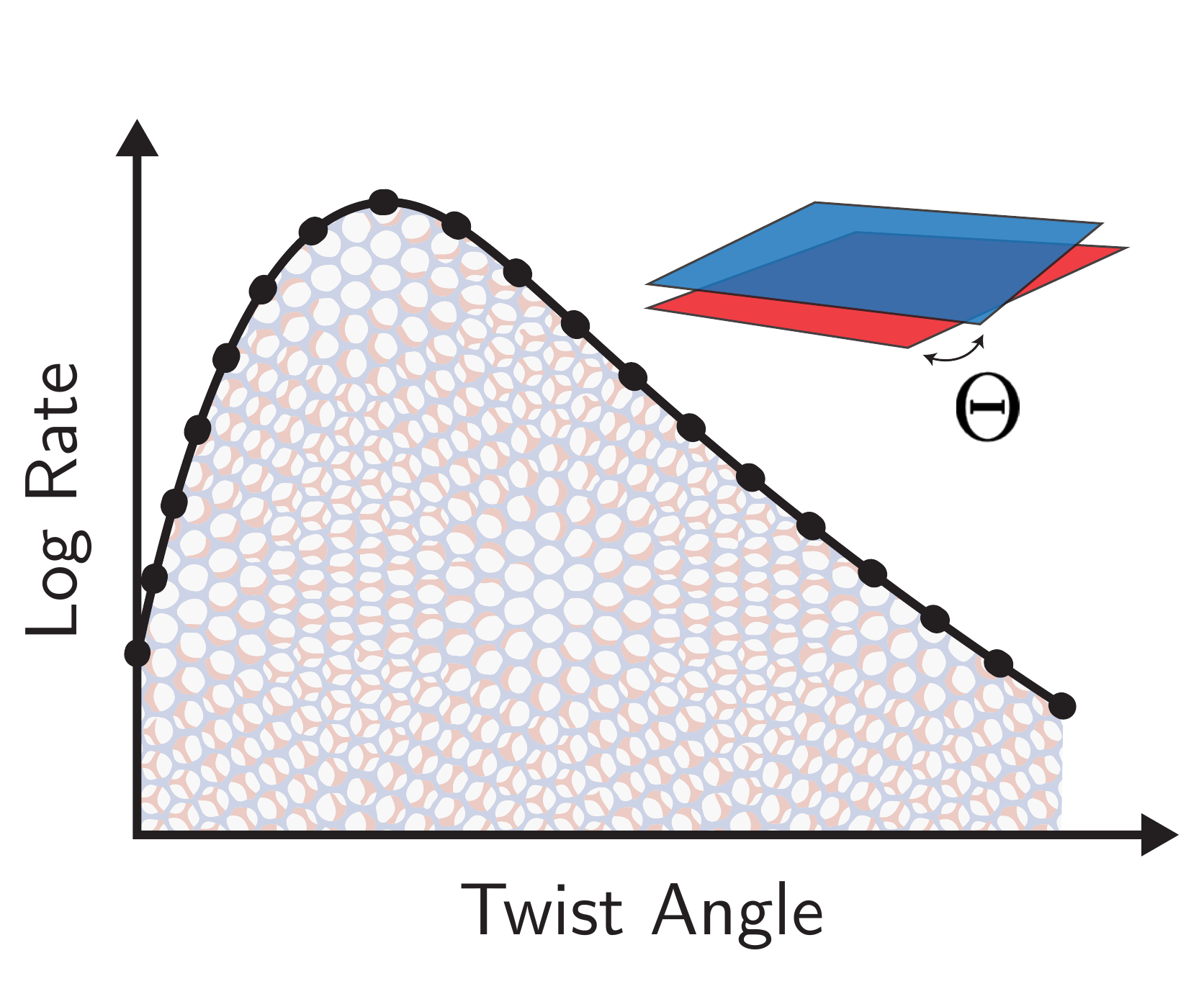}
    \captionsetup{labelformat=empty}
    \caption{TOC}
    \addtocounter{figure}{-1}
\end{figure}
The electron transfer rate measured in Ref. \cite{yu_tunable_2022} was found to have a non-monotonic dependence on twist angle, and was maximized in the vicinity of TBG's first magic angle, $\sim 1.1^\circ$, increasing by nearly two orders of magnitude over its value for an AB Bernal stacked bilayer. The flat bands in magic angle-TBG manifest in the electronic density of states as sharply peaked features or van Hove singularities \cite{bistritzer_moire_2011}. Accounting only for the direct contribution from the increased density of states to the rate within Marcus-Gerischer theory, however, significantly underestimates the extent of the rate acceleration compared to experiment. A possibility that has been so far ignored is an additional angle-dependence in the energetic barrier.
In this work we find that this is indeed the case, and identify angle-dependent image charge interactions at the electrochemical interface as the molecular mechanism that gives rise to this phenomenon, thus presenting a more complete, microscopic picture of the electrochemistry in TBG that is physically intuitive and can be broadly applied to other materials with analogous electronic tunability.

We focus on electron transfer reactions that are well described by Marcus theory \cite{marcus_theory_1956, marcus1957theory, marcus_theory_1965-1}. This is the appropriate regime given that outer-sphere oxidation-reduction (redox) reactions, such as \ch{[Ru(NH_3)_6]^{2+}/[Ru(NH_3)_6]^{3+}}, have been used to probe the electrochemistry of TBG experimentally\cite{yu_tunable_2022}. Outer sphere redox reactions are mediated predominantly by long-ranged electrostatic interactions, where the thermal fluctuations of the polar environment play a central role\cite{chandler_chapter_nodate}. Electron transfer is treated as a nonadiabatic transition between weakly coupled donor ($D$) and acceptor ($A$) states, in which the nuclear degrees of freedom can be approximated as classical. Nonadiabatic processes involve a change in the electronic state of the system at minimally perturbed nuclear configurations, and their rate can be described with time-dependent perturbation theory for small electronic couplings. For interfacial electron transfer into a continuum of electronic states,  the resulting golden rule rate expression is\cite{nitzan_chemical_2014}
\begin{equation}
    \label{interfacialFGR}
        k =  \frac{2\pi}{\hbar} |V_{DA}|^2 \int dE \,D(E) f(E) \langle\delta (\Delta E - E)\rangle_D  \, ,
\end{equation}
where $V_{DA}$ is the diabatic coupling between the electron transfer states, $D(E)$ is the density of electronic states in the electrode, and $f(E)$ is the Fermi-Dirac distribution. The instantaneous, vertical energy gap $\Delta E$  between the two electron transfer states, is defined as $ \Delta E(\textbf{r}^N) = \mathcal{H}_A(\textbf{r}^N) - \mathcal{H}_D(\textbf{r}^N)$, where $\mathcal{H}_i(\textbf{r}^N)$ is the diabatic state Hamiltonian for state $i=\{D,A\}$ at fixed nuclear coordinates but relaxed electronic degrees of freedom. 
The thermal distribution of energy gaps, ${p(\Delta E)}= \langle\delta (\Delta E)\rangle_D$, sampled from the donor Hamiltonian, can be evaluated from molecular simulation. Within linear response theory, it is Gaussian \cite{ferrario_computer_2006, chandler_chapter_nodate}
\begin{equation}
    \label{Marcus_parabola}
    \ln{p(\Delta E)} = - \frac{(\Delta E - \Delta F - \lambda)^2}{4 \kB T \lambda} +\frac{1}{2}\ln [ 4 \pi  \kB T  \lambda ]
\end{equation}
where $\kB T$ is Boltzmann's constant times temperature and $\Delta F$ is the thermodynamic driving force of the reaction. The reorganization energy, $\lambda$, is the work associated with transforming the nuclear configuration of a given charge transfer state at equilibrium, to the equilibrium nuclear configuration of the complementary charge transfer state.

Manipulating the twist angle in TBG can take the material through a variety of electronic phases, from insulating to conducting \cite{wong_cascade_2020, cao_correlated_2018, cao_unconventional_2018, yankowitz_tuning_2019}. This necessarily manifests in changes to the effective electrostatic interactions between charge carriers, meaning that the dielectric response function in TBG depends on twist angle. The simplest model to describe screening between weakly interacting electrons is Thomas-Fermi theory\cite{ashcroft_solid_1976}. This is a semi-classical, mean-field approximation, where electronic screening is prescribed by a dielectric response function with Fourier representation $\hat{\epsilon}(\textbf{k}) = \epsilon^{(\text{el})}\left(1+1/\ltf^2k^2\right)$, where $\epsilon^{(\text{el})}$ is the static dielectric constant of the underlying electrode lattice, and $\ltf$ is the Thomas-Fermi screening length. The screening length is the central parameter of the theory, it sets the characteristic decay length of electrostatic interactions, and it interpolates between a perfect metal ($\ltf=0$), and an insulator ($\ltf\to\infty$). For electron charge, $e$, it is defined as
\begin{equation}
    \label{lTF_defn_full}
    \ltf(\theta) = \sqrt{\frac{\epsilon^{(\text{el})}}{4\pi e^2 \left(\partial\rho/\partial \mu\right)}} \approx  \sqrt{\frac{\epsilon^{(\text{el})}}{4\pi e^2 D(\mu ; \theta)}} \, ,
\end{equation}
where $\rho$ and $\mu$ are the electron density and chemical potential, respectively. The second equality is only exact at zero temperature, and holds as a good approximation provided that $\mu >> \kB T$. Since the density of states depends strongly on the twist angle, $D(\mu;\theta)$, to investigate the effect of twist angle on electron transfer, we propose working with the Thomas-Fermi screening length as an effective proxy for it. This is done within the context of three dimensional electrostatics, but analogous relationships hold in two dimensions\cite{noori_dielectric_2019,ando_electronic_1982,SI}. 

\begin{figure}[t]
    \includegraphics[width=8.5cm]{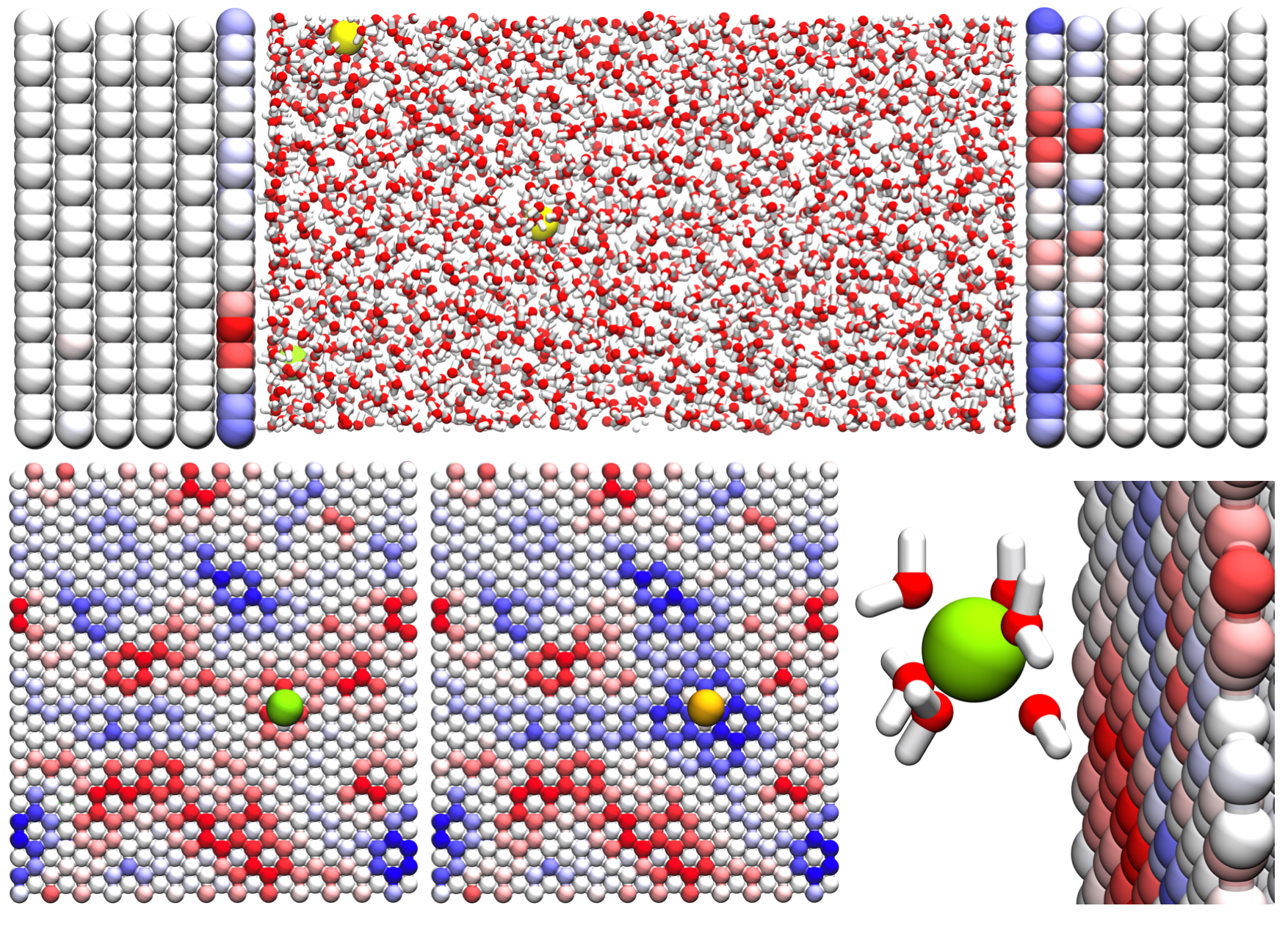}
    \caption{{Top:} Snapshot of the full simulation cell. The green particle is an $\text{Fe}^{2+}$ ion, held a fixed separation in the $z$-direction from the surface of the electrode. Yellow particles are $\text{Cl}^{-}$ counter ions. A color mapping is used to illustrate the value of the fluctuating electrode charges, the scale ranges from $-0.01 e$ (blue) to $+0.01 e$ (red). {Bottom left:} Illustration of the polarization response of the electrode, generated upon switching the charge state of the redox ion (left: $\text{Fe}^{2+}$ at an equilibrium configuration, right: $\text{Fe}^{3+}$ in the same configuration). {Bottom right:} First solvation shell of the iron species, showing an octahedral coordination environment.}
    \label{fig:sim_snapshots}
\end{figure}

Electrochemical reactions in the Marcus regime can be studied with classical simulations \cite{warshel_dynamics_1982, reed_electrochemical_2007, reed_electrochemical_2008, willard_water_2008}. To adequately describe electrochemical charge transfer, we need to include degrees of freedom that represent the fluctuating charge density of the electrode. This is done by working in the constant potential ensemble \cite{siepmann1995influence,scalfi_charge_2020, limmer2013charge, scalfi_molecular_2021}, in which the charges of the electrode atoms evolve throughout the simulation, subject to a constraint of constant potential at the surface of the electrode. A generalization of the constant potential method based on Thomas-Fermi theory has been recently developed to capture deviations from ideal electrode metallicity \cite{scalfi_semiclassical_2020, scalfi_microscopic_2021}. In studying electrochemical charge transfer, the main goal becomes sampling and reconstructing $p(\Delta E)$ from the simulation. We achieve this through thermodynamic integration \cite{hwang_microscopic_1987, king_investigation_1990}, by running simulations and sampling the gap in biased ensembles, governed by Hamiltonian, $\mathcal{H}_\eta  = \mathcal{H}_D + \eta \Delta E$. Where the biasing parameter $\eta$ ranges from 0 to 1. The free energy surfaces were then reconstructed using the multi-state Bennett acceptance ratio algorithm \cite{shirts2008statistically}. 

\begin{figure}[t]
            \includegraphics[width=8.5cm]{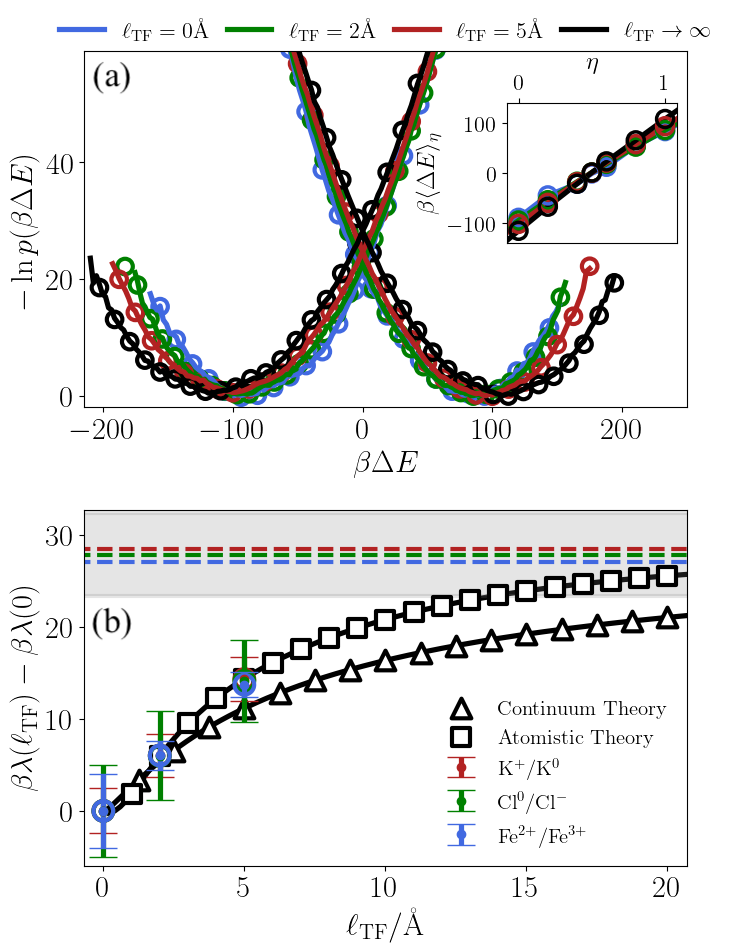}          \caption{(a) Free energy surfaces of electron transfer for $\ch{Fe^3+}/\ch{Fe^2+}$ at different $\ltf$. (b) Reorganization energy for the three redox couples (symbols), as a function of $\ltf$ and comparisons to the energy gap computed electrostatically from the image charge interaction from simulations (squares) and dielectric theory (triangles). Dashed lines are the results from calculations with $\ltf \rightarrow \infty$ and the grey region their statistical errorbar.}
    \label{fig:Marcus_curves}
\end{figure}

Constant potential molecular dynamics simulations were run in LAMMPS,\cite{thompson_lammps_2022,ahrens-iwers_electrode_2022} and energy gap statistics were sampled via thermodynamic integration for three model outer sphere redox couples ($\ch{Fe^3+}/\ch{Fe^2+}$, $\ch{K^+}/\ch{K^0}$ and $\ch{Cl^0}/\ch{Cl^-}$), and four different values of screening length, at zero applied potential and constant temperature $T=298 $K. A snapshot of the simulation is shown in the upper panel of Fig. \ref{fig:sim_snapshots} in which atomistic models of graphitic electrodes enclose an aqueous electrolyte solution.
Approximately 2,300 water molecules interacting through the SPC/E potential  were included in a cell of dimensions $31.5 \mathrm{\AA} \times 34.5 \mathrm{\AA} \times 70 \mathrm{\AA}$, with periodic boundary conditions in the $x$ and $y$ directions, and enclosed on either side of the $z$-axis by six graphitic electrode layers.
 The SPC/E water model was chosen both for its simplicity and its success in reproducing the dielectric properties of water\cite{berendsen_missing_1987, rami_reddy_dielectric_1989}. Simulations were evolved using the finite field method, where the potential difference in the simulation is achieved by applying a uniform electric field across a single electrode with periodic boundary conditions across it, allowing Coulomb interactions to be computed using the regular particle-mesh approach to Ewald summation\cite{dufils_simulating_2019}. Two chloride counter ions were included in the $\ch{Fe^3+}/\ch{Fe^2+}$ system to ensure neutrality in the reduced state. Simple dispersion interactions were modeled with the Lennard-Jones potential\cite{li_rational_2013, hummer_water_2001, yagasaki_lennard-jones_2020} with arithmetic mixing rules used for all species except for the $\ch{Fe^3+}$ and $\ch{Fe^2+}$ ions, which employ a force field that approximately captures their octahedral coordination in aqueous solution \cite{curtiss_nonadditivity_1986, hutchinson_ionic_1999}, depicted on the right of the bottom panel of Fig. \ref{fig:sim_snapshots}. The iron force fields have been used previously in simulation studies of $\ch{Fe^3+}/\ch{Fe^2+}$ electrochemical charge transfer \cite{smith_simulation_1994}. Outer sphere reactions are not affected by the specific molecular details of the redox species, only its charge and size. These three redox systems were chosen because they span a representative range of both.

The pictures on the bottom left of Fig. \ref{fig:sim_snapshots} illustrate how the electrode charges respond to an instantaneous change in charge of the redox center (by $+1 e$), showing an appreciable local accumulation of negative charge in the vicinity of the redox species. This ``image charge" is the electrode's response to the newly formed positive charge on the redox ion, before the molecular dipoles of the solvent reorient to stabilize the redox species. It can be interpreted as the manifestation of an effective charge transfer between the redox center and the electrode. As one increases the screening length in the electrode, however, one effectively reduces its metallic character, limiting its ability to locally polarize. Consequently, at larger values of the screening length, the magnitude and the degree of localization of the image charge will decrease. 

For all of the results presented here, the redox ion was held at an average separation of 5 $\mathrm{\AA}$ from the surface of the electrode, far enough to retain its first solvation shell, and coincident with a local minimum in the potential of mean force. Example electron transfer free energy surfaces for the $\ch{Fe^3+}/\ch{Fe^2+}$ system at all screening lengths are shown in Fig. \ref{fig:Marcus_curves}(a). The free energy surfaces, $-\ln{p(\Delta E)}$, shown in Fig. \ref{fig:Marcus_curves} were all displaced by a constant in order for their intersection to be located at $\Delta E = 0$, to account for the unknown work function of the electrode\cite{reed_electrochemical_2008, willard_water_2008}. Linear response is observed to be valid in all cases. This is confirmed through various metrics, such as the consistency in distinct definitions of the reorganization energy \cite{ferrario_computer_2006, SI}, and the nearly linear relationship of the expectation value of the energy gap in the biased ensembles $\langle\Delta E\rangle_\eta$ as a function of biasing parameter $\eta$ shown in the inset of Fig. \ref{fig:Marcus_curves}(a), where the subscript denotes an average with Hamiltonian, $\mathcal{H}_\eta$.

The simulation results show that electrode screening has a significant effect on the rate of electron transfer, with higher values of screening length leading to a monotonic increase in the free energy of activation, corresponding to the crossing point of the two free energy surfaces in Fig. \ref{fig:Marcus_curves}(a). The trend holds for all redox pairs considered, independent of the identity of the redox species\cite{SI}. For simplicity we work at conditions where the redox process is isoenergetic, $\Delta F=0$, so the barrier height from Marcus theory reduces to just the reorganization energy, $\lambda/4$. Indeed, the changes in the curvature and the relative separation between the minima confirm that the reorganization energy is an increasing function of screening length. Shown in Fig.~\ref{fig:Marcus_curves}(b) are the changes to the reorganization energy as a function of screening length for all three redox pairs, which we find agree with each other quantitatively. This is reasonable, because we expect the effect of screening to be purely electrostatic in nature, so the square of the change in charge upon vertical excitation is the only property that effectively distinguishes one redox system from another and for all cases considered this is equal to 1$e$. Furthermore, this generic behavior strongly suggests that our findings are broadly applicable to redox reactions that proceed through an outer sphere mechanism, such as the \ch{[Ru(NH_3)_6] ^{2+/3+}} species used in the experimental study. The small differences that we do observe between redox systems may be attributed to deviations from ideal dielectric theory, as it is well-known that water solvates ions of opposite sign differently due to its finite quadrupole moment\cite{cox_quadrupole-mediated_2021,cox2022dielectric} or finite size effects\cite{nair2024induced}.

We find that image charge interactions play a central role in determining the behavior of $\lambda(\ltf)$, and they can be described remarkably well with a simple dielectric continuum theory \cite{liu1994reorganization}. A full approach would need to contend with the anisotropic dielectric response inherent to the electrochemical interface, a framework which has been developed and studied at length, most prominently by Kornyshev and co-workers \cite{medvedev_nonlocal_2002, dzhavakhidze_activation_1987, medvedev_non-local_2000, a_kornyshev__1977}. In the following, however, we will rely on a simpler treatment\cite{fedorov_ionic_2014,kaiser_electrostatic_2017}, which considers the electrode and electrolyte to be well described by their bulk properties up to a sharp boundary between them. 

For a point charge $q$ in a solution with dielectric constant $\epsilon^{(\text{sol})}$ located a distance $z_0>0$ away from a planar interface with an electrode characterized by screening length $\ltf$ and a static dielectric constant $\epsilon^{(\text{el})}$, Poisson's equation at the boundary may be solved making use of Fourier-Bessel transforms. Although the resulting electrostatic potential is not analytically invertible, an expression for the electric potential energy $U$ between the charge and the electrode may still be derived, given by
\begin{equation}
    \label{Uim}
    U(z_0,\ltf) = \frac{q^2\xi_{\ltf}(z_0,\epsilon^{(\text{sol})})}{4\epsilon^{(\text{sol})}z_0} 
\end{equation}
where we define an image charge scaling function, $\xi_{\ltf}(z_0,\epsilon^{(\text{sol})})$, as
\begin{equation}
    \xi_{\ltf}(z_0,\epsilon^{(\text{sol})}) = \mathcal{I}_1 -\epsilon^{(\text{sol})}\mathcal{I}_2 -1
\end{equation}
introducing the dimensionless parameter $\zeta \equiv z_0/ \ltf$, the integrals,
\begin{equation}
    \label{I1}
    \mathcal{I}_1(\zeta) = \int_0^\infty d\tau \frac{4\left(\frac{\epsilon^{(\text{sol})}}{\epsilon^{(\text{el})}}\right)\tau e^{-2\tau}}{\sqrt{\tau^2+\zeta^2}+\left(\frac{\epsilon^{(\text{sol})}}{\epsilon^{(\text{el})}}\right)\tau}
\end{equation}
and
\begin{equation}
    \label{I2}
    \mathcal{I}_2(\zeta) = \int_0^\infty d\tau \frac{2 \zeta^2\tau e^{-2\tau}}{\left(\sqrt{\tau^2+\zeta^2}+\left(\frac{\epsilon^{(\text{sol})}}{\epsilon^{(\text{el})}}\right)\tau\right)^2\sqrt{\tau^2+\zeta^2}}
\end{equation}
are related to the inverse Fourier-Bessel transforms of the potential and the charge density in the electrode. The appearance of Eq. \ref{Uim} suggests a simple and intuitive interpretation, as the Coulomb energy between the charge and a fictitious image point charge located at $-z_0$ whose magnitude depends on $z_0$,  $\ltf$, and the dielectric constants of the media. The potential smoothly interpolates between the electrostatics of a charge at the boundary of an ideal conductor for $\ltf=0$, $ \xi_{0}(z_0,\epsilon^{(\text{sol})}) = -1$ and an insulator in the limit $\ltf\to\infty$,  $ \xi_{\infty}(z_0,\epsilon^{(\text{sol})}) = (\epsilon^{(\text{sol})}-1)/(\epsilon^{(\text{sol})}+1)$, with $\epsilon^{(\text{el})} = 1$.

In Marcus theory, the expectation value of the energy gap is completely determined by the reorganization energy and the driving force of the reaction, and for $\Delta F=0$, $\langle\Delta E\rangle_D = \lambda$. Accounting exclusively for electrostatic interactions within dielectric continuum theory, the vertical energy gap, and therefore the contribution to the reorganization energy from the electrode, can be computed from $U(z_0,\ltf)$. If we consider the interaction between the instantaneously formed excess charge $\delta q$ and its image in the electrode to only be shielded by the fast electronic degrees of freedom of the solvent, characterized by the optical dielectric constant $\epsilon_{\infty}^{(\text{sol})}$, the reorganization energy can be expressed as
\begin{equation}
    \label{lambda_ellTF}
    \begin{split}
        \lambda =& \frac{\delta q^2 }{4z_0} \left(\frac{\xi_{\ltf}(z_0,\epsilon^{\text{(sol)}}_{\infty})}{\epsilon^{\text{(sol)}}_{\infty}}-\frac{\xi_{\ltf}(z_0,\epsilon^{\text{(sol)}})}{\epsilon^{\text{(sol)}}}\right) + \lambda_\mathrm{B}
            \end{split}
\end{equation}
where $\lambda_\mathrm{B}$ corresponds to a pure solvent contribution to the reorganization energy that is independent of screening in the electrode within the sharp-boundary approximation\cite{SI}. The term inside parenthesis in Eq. \ref{lambda_ellTF} can be recognized as a generalization of the usual Pekar factor that features in continuum theories of solvation and charge transfer, extended to describe solvation of charges close to the Thomas-Fermi electrode. 

Equation ~\ref{lambda_ellTF} reduces to the well-known estimate from Marcus for electrode reactions, for $\ltf=0$, if we take $\lambda_\mathrm{B}$ as Marcus' dielectric continuum estimate for the reorganization energy of bulk reactions\cite{limaye2020understanding}. The function $\xi_{\ltf}$ may be computed numerically for a perfectly planar electrode by evaluating Eqs. \ref{I1} and \ref{I2}, but it can also be obtained for an atomistic electrode by measuring the potential energy  for a test charge in the simulation without solvent. Figure \ref{fig:Marcus_curves}(b) shows a comparison of $\lambda(\ltf)$ obtained from simulation, and two predictions derived from Eq. ~\ref{lambda_ellTF}. One is the pure continuum result, while the other is a theoretical prediction obtained by constructing $\xi_{\ltf}(z_0,\epsilon_\infty^{\mathrm{(sol)}})$ using the potential from the atomistic charge density. There is nearly exact agreement between the latter and simulation data, while the pure continuum version shows the correct qualitative trend and is in good agreement at low values of $\ltf$, but it deviates at larger ones. This discrepancy is a consequence of the atomistic nature of the electrode, whose induced charge distribution exhibits many-body interactions with a charge placed very close to it, which are not considered in a continuum description \cite{nair2024induced}.

\begin{figure}[t]
            \includegraphics[width=8.5cm]{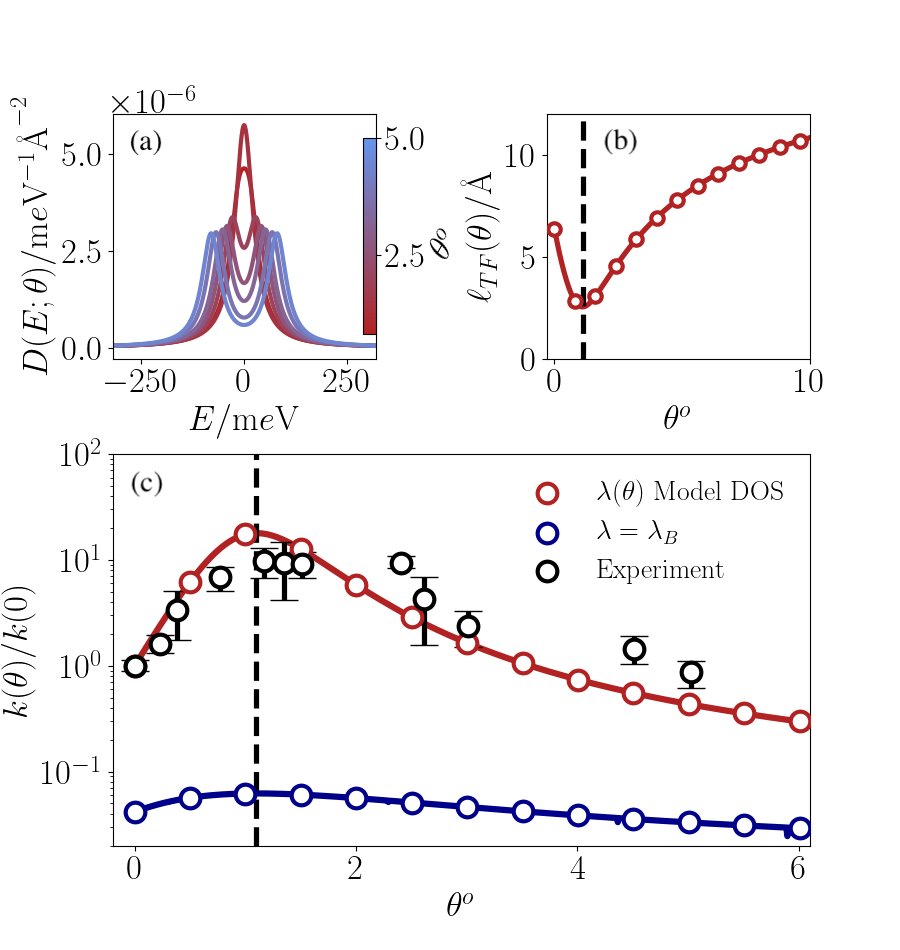}
    \caption{(a) Density of states for various twist angles for an empirical double-Lorentzian model. (b) Screening length computed from $D(E)$ in (a), for $E=0$ as a function of $\theta$. (c) Rate as function of twist angle evaluated using $D(E)$ and $\ltf$ in (a) and (b) (red line), compared to the rate obtained with fixed $\lambda$ (blue line), and the experimental measurements of the rate \cite{yu_tunable_2022}.} Dashed black line indicates the magic angle, 1.1$^\circ$
    \label{fig:rate_angle}
\end{figure}

With an understanding for how $p(\Delta E)$ changes with $\ltf$, and therefore $\theta$, we can evaluate the integral in Eq. \ref{interfacialFGR} to construct a twist-dependent rate. To do so, we have employed a simple model density of states, $D(E;\theta)$, shown in Fig.~\ref{fig:rate_angle}(a) constructed to reproduce calculations of TBG in Ref. \cite{yu_tunable_2022} near the Fermi level\cite{SI}. This determines an angle-dependent $\ltf$ shown in Fig.~\ref{fig:rate_angle}(b), which is minimized at the magic angle due to the emergent flat bands and increased density of states. The behavior of the rate of interfacial electron transfer using this model is shown in Fig. \ref{fig:rate_angle} (c), alongside the experimental rate measurements\cite{yu_tunable_2022}. Shown is the angle-dependent rate relative to the rate in a Bernal stacked configuration of $\theta=0^{\circ}$. Accounting for changes in the reorganization energy leads to a dramatic rate enhancement at low angles. We compare the rate obtained from an angle-dependent reorganization energy with the rate that results from considering a constant reorganization energy, $\lambda=\lambda_\mathrm{B}$. The full angle-dependent density of states was used in both cases, so they only differ in how the reorganization energy is being treated. The curves exhibit a peak at the magic angle. In the case of fixed $\lambda$, this modest increase can be associated purely to the contribution from the density of states, which enters into the rate expression linearly, while the exponential effect of a maximally attenuated reorganization energy at the magic angle leads to a rate that is faster by over an of magnitude, in agreement with experimental data. This behavior is robust to different models of the density of states. Modeling the effects of lattice relaxation results in a broader peak in the rate that is flatter at small twist angles\cite{SI}. This is consistent with the AA site-specific rate measurements.

By mapping the effect of twist angle in TBG to the non-local dielectric response of the electrode, we have proposed a microscopic origin for the observed electron transfer rate enhancement. The effect of non-local electrostatic dispersion on the rate of electron transfer has been a subject of previous study \cite{dzhavakhidze_activation_1987, medvedev_nonlocal_2002}, but here we have shown that stacked two-dimensional materials offer a potential platform for exploring it. Further, leveraging recent advances in molecular simulation we have provided a link between continuum and atomistic perspectives, clarifying the physical interpretation of the former in the process. In our treatment, we have ignored the effect that the angle of twist may have on the diabatic coupling factor $V_{DA}$. We have also ignored the in-plane heterogeneity of the Moiré superlattice that breaks the surface of TBG into domains of higher and lower local density of states with distinct local stacking environments. Both could be included straightforwardly. Nevertheless, we find that accounting for the twist-dependent reorganization energy that results from the altered ability of the electrode to produce a stabilizing image charge has allowed us to bridge the gap between experiment and theory in the electrochemistry of twisted bilayer graphene.

The source code for an example simulation and the data that supports the findings of this study are openly
available online \cite{SuppData}. Supporting Information: additional details on the simulations and the construction of Marcus curves, derivation of the screening-dependent reorganization energy, comparison of the rate calculations presented here to alternative models for the density of states, and to energetically misaligned redox couples.

\emph{Awknowledgements.} We would like to thank Kwabena Bediako for motivating this study, and Seokjin Moon, Benjamin Rotenberg, and Mathieu Salanne for helpful discussions regarding the simulations. L.C.E. and D.T.L. were supported by the U.S. Department of Energy, Office of Science, Basic Energy Sciences, CPIMS Program Early Career Research Program under Award DEFOA0002019.

\section*{References}

\bibliography{references,ref2}

\begin{thebibliography}{66}%
\makeatletter
\providecommand \@ifxundefined [1]{%
 \@ifx{#1\undefined}
}%
\providecommand \@ifnum [1]{%
 \ifnum #1\expandafter \@firstoftwo
 \else \expandafter \@secondoftwo
 \fi
}%
\providecommand \@ifx [1]{%
 \ifx #1\expandafter \@firstoftwo
 \else \expandafter \@secondoftwo
 \fi
}%
\providecommand \natexlab [1]{#1}%
\providecommand \enquote  [1]{``#1''}%
\providecommand \bibnamefont  [1]{#1}%
\providecommand \bibfnamefont [1]{#1}%
\providecommand \citenamefont [1]{#1}%
\providecommand \href@noop [0]{\@secondoftwo}%
\providecommand \href [0]{\begingroup \@sanitize@url \@href}%
\providecommand \@href[1]{\@@startlink{#1}\@@href}%
\providecommand \@@href[1]{\endgroup#1\@@endlink}%
\providecommand \@sanitize@url [0]{\catcode `\\12\catcode `\$12\catcode
  `\&12\catcode `\#12\catcode `\^12\catcode `\_12\catcode `\%12\relax}%
\providecommand \@@startlink[1]{}%
\providecommand \@@endlink[0]{}%
\providecommand \url  [0]{\begingroup\@sanitize@url \@url }%
\providecommand \@url [1]{\endgroup\@href {#1}{\urlprefix }}%
\providecommand \urlprefix  [0]{URL }%
\providecommand \Eprint [0]{\href }%
\providecommand \doibase [0]{https://doi.org/}%
\providecommand \selectlanguage [0]{\@gobble}%
\providecommand \bibinfo  [0]{\@secondoftwo}%
\providecommand \bibfield  [0]{\@secondoftwo}%
\providecommand \translation [1]{[#1]}%
\providecommand \BibitemOpen [0]{}%
\providecommand \bibitemStop [0]{}%
\providecommand \bibitemNoStop [0]{.\EOS\space}%
\providecommand \EOS [0]{\spacefactor3000\relax}%
\providecommand \BibitemShut  [1]{\csname bibitem#1\endcsname}%
\let\auto@bib@innerbib\@empty
\bibitem [{\citenamefont {Fiori}\ \emph {et~al.}(2014)\citenamefont {Fiori},
  \citenamefont {Bonaccorso}, \citenamefont {Iannaccone}, \citenamefont
  {Palacios}, \citenamefont {Neumaier}, \citenamefont {Seabaugh}, \citenamefont
  {Banerjee},\ and\ \citenamefont {Colombo}}]{fiori_electronics_2014}%
  \BibitemOpen
  \bibfield  {author} {\bibinfo {author} {\bibfnamefont {G.}~\bibnamefont
  {Fiori}}, \bibinfo {author} {\bibfnamefont {F.}~\bibnamefont {Bonaccorso}},
  \bibinfo {author} {\bibfnamefont {G.}~\bibnamefont {Iannaccone}}, \bibinfo
  {author} {\bibfnamefont {T.}~\bibnamefont {Palacios}}, \bibinfo {author}
  {\bibfnamefont {D.}~\bibnamefont {Neumaier}}, \bibinfo {author}
  {\bibfnamefont {A.}~\bibnamefont {Seabaugh}}, \bibinfo {author}
  {\bibfnamefont {S.~K.}\ \bibnamefont {Banerjee}},\ and\ \bibinfo {author}
  {\bibfnamefont {L.}~\bibnamefont {Colombo}},\ }\bibfield  {title} {\bibinfo
  {title} {Electronics based on two-dimensional materials},\ }\href
  {https://doi.org/10.1038/nnano.2014.207} {\bibfield  {journal} {\bibinfo
  {journal} {Nature Nanotechnology}\ }\textbf {\bibinfo {volume} {9}},\
  \bibinfo {pages} {768} (\bibinfo {year} {2014})}\BibitemShut {NoStop}%
\bibitem [{\citenamefont {Jin}\ \emph {et~al.}(2018)\citenamefont {Jin},
  \citenamefont {Guo}, \citenamefont {Liu}, \citenamefont {Liu}, \citenamefont
  {Vasileff}, \citenamefont {Jiao}, \citenamefont {Zheng},\ and\ \citenamefont
  {Qiao}}]{jin_emerging_2018}%
  \BibitemOpen
  \bibfield  {author} {\bibinfo {author} {\bibfnamefont {H.}~\bibnamefont
  {Jin}}, \bibinfo {author} {\bibfnamefont {C.}~\bibnamefont {Guo}}, \bibinfo
  {author} {\bibfnamefont {X.}~\bibnamefont {Liu}}, \bibinfo {author}
  {\bibfnamefont {J.}~\bibnamefont {Liu}}, \bibinfo {author} {\bibfnamefont
  {A.}~\bibnamefont {Vasileff}}, \bibinfo {author} {\bibfnamefont
  {Y.}~\bibnamefont {Jiao}}, \bibinfo {author} {\bibfnamefont {Y.}~\bibnamefont
  {Zheng}},\ and\ \bibinfo {author} {\bibfnamefont {S.-Z.}\ \bibnamefont
  {Qiao}},\ }\bibfield  {title} {\bibinfo {title} {Emerging {Two}-{Dimensional}
  {Nanomaterials} for {Electrocatalysis}},\ }\href
  {https://doi.org/10.1021/acs.chemrev.7b00689} {\bibfield  {journal} {\bibinfo
   {journal} {Chemical Reviews}\ }\textbf {\bibinfo {volume} {118}},\ \bibinfo
  {pages} {6337} (\bibinfo {year} {2018})}\BibitemShut {NoStop}%
\bibitem [{\citenamefont {Chia}\ and\ \citenamefont
  {Pumera}(2018)}]{chia_characteristics_2018}%
  \BibitemOpen
  \bibfield  {author} {\bibinfo {author} {\bibfnamefont {X.}~\bibnamefont
  {Chia}}\ and\ \bibinfo {author} {\bibfnamefont {M.}~\bibnamefont {Pumera}},\
  }\bibfield  {title} {\bibinfo {title} {Characteristics and performance of
  two-dimensional materials for electrocatalysis},\ }\href
  {https://doi.org/10.1038/s41929-018-0181-7} {\bibfield  {journal} {\bibinfo
  {journal} {Nature Catalysis}\ }\textbf {\bibinfo {volume} {1}},\ \bibinfo
  {pages} {909} (\bibinfo {year} {2018})}\BibitemShut {NoStop}%
\bibitem [{\citenamefont {Geim}\ and\ \citenamefont
  {Grigorieva}(2013)}]{geim_van_2013}%
  \BibitemOpen
  \bibfield  {author} {\bibinfo {author} {\bibfnamefont {A.~K.}\ \bibnamefont
  {Geim}}\ and\ \bibinfo {author} {\bibfnamefont {I.~V.}\ \bibnamefont
  {Grigorieva}},\ }\bibfield  {title} {\bibinfo {title} {Van der {Waals}
  heterostructures},\ }\href {https://doi.org/10.1038/nature12385} {\bibfield
  {journal} {\bibinfo  {journal} {Nature}\ }\textbf {\bibinfo {volume} {499}},\
  \bibinfo {pages} {419} (\bibinfo {year} {2013})}\BibitemShut {NoStop}%
\bibitem [{\citenamefont {Andrei}\ \emph {et~al.}(2021)\citenamefont {Andrei},
  \citenamefont {Efetov}, \citenamefont {Jarillo-Herrero}, \citenamefont
  {MacDonald}, \citenamefont {Mak}, \citenamefont {Senthil}, \citenamefont
  {Tutuc}, \citenamefont {Yazdani},\ and\ \citenamefont
  {Young}}]{andrei_marvels_2021}%
  \BibitemOpen
  \bibfield  {author} {\bibinfo {author} {\bibfnamefont {E.~Y.}\ \bibnamefont
  {Andrei}}, \bibinfo {author} {\bibfnamefont {D.~K.}\ \bibnamefont {Efetov}},
  \bibinfo {author} {\bibfnamefont {P.}~\bibnamefont {Jarillo-Herrero}},
  \bibinfo {author} {\bibfnamefont {A.~H.}\ \bibnamefont {MacDonald}}, \bibinfo
  {author} {\bibfnamefont {K.~F.}\ \bibnamefont {Mak}}, \bibinfo {author}
  {\bibfnamefont {T.}~\bibnamefont {Senthil}}, \bibinfo {author} {\bibfnamefont
  {E.}~\bibnamefont {Tutuc}}, \bibinfo {author} {\bibfnamefont
  {A.}~\bibnamefont {Yazdani}},\ and\ \bibinfo {author} {\bibfnamefont {A.~F.}\
  \bibnamefont {Young}},\ }\bibfield  {title} {\bibinfo {title} {The marvels of
  moiré materials},\ }\href {https://doi.org/10.1038/s41578-021-00284-1}
  {\bibfield  {journal} {\bibinfo  {journal} {Nature Reviews Materials}\
  }\textbf {\bibinfo {volume} {6}},\ \bibinfo {pages} {201} (\bibinfo {year}
  {2021})}\BibitemShut {NoStop}%
\bibitem [{\citenamefont {He}\ \emph {et~al.}(2021)\citenamefont {He},
  \citenamefont {Zhou}, \citenamefont {Ye}, \citenamefont {Cho}, \citenamefont
  {Jeong}, \citenamefont {Meng},\ and\ \citenamefont {Wang}}]{he_moire_2021}%
  \BibitemOpen
  \bibfield  {author} {\bibinfo {author} {\bibfnamefont {F.}~\bibnamefont
  {He}}, \bibinfo {author} {\bibfnamefont {Y.}~\bibnamefont {Zhou}}, \bibinfo
  {author} {\bibfnamefont {Z.}~\bibnamefont {Ye}}, \bibinfo {author}
  {\bibfnamefont {S.-H.}\ \bibnamefont {Cho}}, \bibinfo {author} {\bibfnamefont
  {J.}~\bibnamefont {Jeong}}, \bibinfo {author} {\bibfnamefont
  {X.}~\bibnamefont {Meng}},\ and\ \bibinfo {author} {\bibfnamefont
  {Y.}~\bibnamefont {Wang}},\ }\bibfield  {title} {\bibinfo {title} {Moiré
  {Patterns} in {2D} {Materials}: {A} {Review}},\ }\href
  {https://doi.org/10.1021/acsnano.0c10435} {\bibfield  {journal} {\bibinfo
  {journal} {ACS Nano}\ }\textbf {\bibinfo {volume} {15}},\ \bibinfo {pages}
  {5944} (\bibinfo {year} {2021})}\BibitemShut {NoStop}%
\bibitem [{\citenamefont {Nimbalkar}\ and\ \citenamefont
  {Kim}(2020)}]{nimbalkar_opportunities_2020}%
  \BibitemOpen
  \bibfield  {author} {\bibinfo {author} {\bibfnamefont {A.}~\bibnamefont
  {Nimbalkar}}\ and\ \bibinfo {author} {\bibfnamefont {H.}~\bibnamefont
  {Kim}},\ }\bibfield  {title} {\bibinfo {title} {Opportunities and
  {Challenges} in {Twisted} {Bilayer} {Graphene}: {A} {Review}},\ }\href
  {https://doi.org/10.1007/s40820-020-00464-8} {\bibfield  {journal} {\bibinfo
  {journal} {Nano-Micro Letters}\ }\textbf {\bibinfo {volume} {12}},\ \bibinfo
  {pages} {126} (\bibinfo {year} {2020})}\BibitemShut {NoStop}%
\bibitem [{\citenamefont {Andrei}\ and\ \citenamefont
  {MacDonald}(2020)}]{andrei_graphene_2020}%
  \BibitemOpen
  \bibfield  {author} {\bibinfo {author} {\bibfnamefont {E.~Y.}\ \bibnamefont
  {Andrei}}\ and\ \bibinfo {author} {\bibfnamefont {A.~H.}\ \bibnamefont
  {MacDonald}},\ }\bibfield  {title} {\bibinfo {title} {Graphene bilayers with
  a twist},\ }\href {https://doi.org/10.1038/s41563-020-00840-0} {\bibfield
  {journal} {\bibinfo  {journal} {Nature Materials}\ }\textbf {\bibinfo
  {volume} {19}},\ \bibinfo {pages} {1265} (\bibinfo {year}
  {2020})}\BibitemShut {NoStop}%
\bibitem [{\citenamefont {Cao}\ \emph {et~al.}(2018{\natexlab{a}})\citenamefont
  {Cao}, \citenamefont {Fatemi}, \citenamefont {Fang}, \citenamefont
  {Watanabe}, \citenamefont {Taniguchi}, \citenamefont {Kaxiras},\ and\
  \citenamefont {Jarillo-Herrero}}]{cao_unconventional_2018}%
  \BibitemOpen
  \bibfield  {author} {\bibinfo {author} {\bibfnamefont {Y.}~\bibnamefont
  {Cao}}, \bibinfo {author} {\bibfnamefont {V.}~\bibnamefont {Fatemi}},
  \bibinfo {author} {\bibfnamefont {S.}~\bibnamefont {Fang}}, \bibinfo {author}
  {\bibfnamefont {K.}~\bibnamefont {Watanabe}}, \bibinfo {author}
  {\bibfnamefont {T.}~\bibnamefont {Taniguchi}}, \bibinfo {author}
  {\bibfnamefont {E.}~\bibnamefont {Kaxiras}},\ and\ \bibinfo {author}
  {\bibfnamefont {P.}~\bibnamefont {Jarillo-Herrero}},\ }\bibfield  {title}
  {\bibinfo {title} {Unconventional superconductivity in magic-angle graphene
  superlattices},\ }\href {https://doi.org/10.1038/nature26160} {\bibfield
  {journal} {\bibinfo  {journal} {Nature}\ }\textbf {\bibinfo {volume} {556}},\
  \bibinfo {pages} {43} (\bibinfo {year} {2018}{\natexlab{a}})}\BibitemShut
  {NoStop}%
\bibitem [{\citenamefont {Cao}\ \emph {et~al.}(2018{\natexlab{b}})\citenamefont
  {Cao}, \citenamefont {Fatemi}, \citenamefont {Demir}, \citenamefont {Fang},
  \citenamefont {Tomarken}, \citenamefont {Luo}, \citenamefont
  {Sanchez-Yamagishi}, \citenamefont {Watanabe}, \citenamefont {Taniguchi},
  \citenamefont {Kaxiras}, \citenamefont {Ashoori},\ and\ \citenamefont
  {Jarillo-Herrero}}]{cao_correlated_2018}%
  \BibitemOpen
  \bibfield  {author} {\bibinfo {author} {\bibfnamefont {Y.}~\bibnamefont
  {Cao}}, \bibinfo {author} {\bibfnamefont {V.}~\bibnamefont {Fatemi}},
  \bibinfo {author} {\bibfnamefont {A.}~\bibnamefont {Demir}}, \bibinfo
  {author} {\bibfnamefont {S.}~\bibnamefont {Fang}}, \bibinfo {author}
  {\bibfnamefont {S.~L.}\ \bibnamefont {Tomarken}}, \bibinfo {author}
  {\bibfnamefont {J.~Y.}\ \bibnamefont {Luo}}, \bibinfo {author} {\bibfnamefont
  {J.~D.}\ \bibnamefont {Sanchez-Yamagishi}}, \bibinfo {author} {\bibfnamefont
  {K.}~\bibnamefont {Watanabe}}, \bibinfo {author} {\bibfnamefont
  {T.}~\bibnamefont {Taniguchi}}, \bibinfo {author} {\bibfnamefont
  {E.}~\bibnamefont {Kaxiras}}, \bibinfo {author} {\bibfnamefont {R.~C.}\
  \bibnamefont {Ashoori}},\ and\ \bibinfo {author} {\bibfnamefont
  {P.}~\bibnamefont {Jarillo-Herrero}},\ }\bibfield  {title} {\bibinfo {title}
  {Correlated insulator behaviour at half-filling in magic-angle graphene
  superlattices},\ }\href {https://doi.org/10.1038/nature26154} {\bibfield
  {journal} {\bibinfo  {journal} {Nature}\ }\textbf {\bibinfo {volume} {556}},\
  \bibinfo {pages} {80} (\bibinfo {year} {2018}{\natexlab{b}})}\BibitemShut
  {NoStop}%
\bibitem [{\citenamefont {Yankowitz}\ \emph {et~al.}(2019)\citenamefont
  {Yankowitz}, \citenamefont {Chen}, \citenamefont {Polshyn}, \citenamefont
  {Zhang}, \citenamefont {Watanabe}, \citenamefont {Taniguchi}, \citenamefont
  {Graf}, \citenamefont {Young},\ and\ \citenamefont
  {Dean}}]{yankowitz_tuning_2019}%
  \BibitemOpen
  \bibfield  {author} {\bibinfo {author} {\bibfnamefont {M.}~\bibnamefont
  {Yankowitz}}, \bibinfo {author} {\bibfnamefont {S.}~\bibnamefont {Chen}},
  \bibinfo {author} {\bibfnamefont {H.}~\bibnamefont {Polshyn}}, \bibinfo
  {author} {\bibfnamefont {Y.}~\bibnamefont {Zhang}}, \bibinfo {author}
  {\bibfnamefont {K.}~\bibnamefont {Watanabe}}, \bibinfo {author}
  {\bibfnamefont {T.}~\bibnamefont {Taniguchi}}, \bibinfo {author}
  {\bibfnamefont {D.}~\bibnamefont {Graf}}, \bibinfo {author} {\bibfnamefont
  {A.~F.}\ \bibnamefont {Young}},\ and\ \bibinfo {author} {\bibfnamefont
  {C.~R.}\ \bibnamefont {Dean}},\ }\bibfield  {title} {\bibinfo {title} {Tuning
  superconductivity in twisted bilayer graphene},\ }\href
  {https://doi.org/10.1126/science.aav1910} {\bibfield  {journal} {\bibinfo
  {journal} {Science}\ }\textbf {\bibinfo {volume} {363}},\ \bibinfo {pages}
  {1059} (\bibinfo {year} {2019})}\BibitemShut {NoStop}%
\bibitem [{\citenamefont {Choi}\ \emph {et~al.}(2019)\citenamefont {Choi},
  \citenamefont {Kemmer}, \citenamefont {Peng}, \citenamefont {Thomson},
  \citenamefont {Arora}, \citenamefont {Polski}, \citenamefont {Zhang},
  \citenamefont {Ren}, \citenamefont {Alicea}, \citenamefont {Refael},
  \citenamefont {von Oppen}, \citenamefont {Watanabe}, \citenamefont
  {Taniguchi},\ and\ \citenamefont {Nadj-Perge}}]{choi_imaging_2019}%
  \BibitemOpen
  \bibfield  {author} {\bibinfo {author} {\bibfnamefont {Y.}~\bibnamefont
  {Choi}}, \bibinfo {author} {\bibfnamefont {J.}~\bibnamefont {Kemmer}},
  \bibinfo {author} {\bibfnamefont {Y.}~\bibnamefont {Peng}}, \bibinfo {author}
  {\bibfnamefont {A.}~\bibnamefont {Thomson}}, \bibinfo {author} {\bibfnamefont
  {H.}~\bibnamefont {Arora}}, \bibinfo {author} {\bibfnamefont
  {R.}~\bibnamefont {Polski}}, \bibinfo {author} {\bibfnamefont
  {Y.}~\bibnamefont {Zhang}}, \bibinfo {author} {\bibfnamefont
  {H.}~\bibnamefont {Ren}}, \bibinfo {author} {\bibfnamefont {J.}~\bibnamefont
  {Alicea}}, \bibinfo {author} {\bibfnamefont {G.}~\bibnamefont {Refael}},
  \bibinfo {author} {\bibfnamefont {F.}~\bibnamefont {von Oppen}}, \bibinfo
  {author} {\bibfnamefont {K.}~\bibnamefont {Watanabe}}, \bibinfo {author}
  {\bibfnamefont {T.}~\bibnamefont {Taniguchi}},\ and\ \bibinfo {author}
  {\bibfnamefont {S.}~\bibnamefont {Nadj-Perge}},\ }\bibfield  {title}
  {\bibinfo {title} {Imaging {Electronic} {Correlations} in {Twisted} {Bilayer}
  {Graphene} near the {Magic} {Angle}},\ }\href
  {https://doi.org/10.1038/s41567-019-0606-5} {\bibfield  {journal} {\bibinfo
  {journal} {Nature Physics}\ }\textbf {\bibinfo {volume} {15}},\ \bibinfo
  {pages} {1174} (\bibinfo {year} {2019})}\BibitemShut {NoStop}%
\bibitem [{\citenamefont {Seh}\ \emph {et~al.}(2017)\citenamefont {Seh},
  \citenamefont {Kibsgaard}, \citenamefont {Dickens}, \citenamefont
  {Chorkendorff}, \citenamefont {Nørskov},\ and\ \citenamefont
  {Jaramillo}}]{seh_combining_2017}%
  \BibitemOpen
  \bibfield  {author} {\bibinfo {author} {\bibfnamefont {Z.~W.}\ \bibnamefont
  {Seh}}, \bibinfo {author} {\bibfnamefont {J.}~\bibnamefont {Kibsgaard}},
  \bibinfo {author} {\bibfnamefont {C.~F.}\ \bibnamefont {Dickens}}, \bibinfo
  {author} {\bibfnamefont {I.}~\bibnamefont {Chorkendorff}}, \bibinfo {author}
  {\bibfnamefont {J.~K.}\ \bibnamefont {Nørskov}},\ and\ \bibinfo {author}
  {\bibfnamefont {T.~F.}\ \bibnamefont {Jaramillo}},\ }\bibfield  {title}
  {\bibinfo {title} {Combining theory and experiment in electrocatalysis:
  {Insights} into materials design},\ }\href
  {https://doi.org/10.1126/science.aad4998} {\bibfield  {journal} {\bibinfo
  {journal} {Science}\ }\textbf {\bibinfo {volume} {355}},\ \bibinfo {pages}
  {eaad4998} (\bibinfo {year} {2017})}\BibitemShut {NoStop}%
\bibitem [{\citenamefont {Stamenkovic}\ \emph {et~al.}(2017)\citenamefont
  {Stamenkovic}, \citenamefont {Strmcnik}, \citenamefont {Lopes},\ and\
  \citenamefont {Markovic}}]{stamenkovic_energy_2017}%
  \BibitemOpen
  \bibfield  {author} {\bibinfo {author} {\bibfnamefont {V.~R.}\ \bibnamefont
  {Stamenkovic}}, \bibinfo {author} {\bibfnamefont {D.}~\bibnamefont
  {Strmcnik}}, \bibinfo {author} {\bibfnamefont {P.~P.}\ \bibnamefont
  {Lopes}},\ and\ \bibinfo {author} {\bibfnamefont {N.~M.}\ \bibnamefont
  {Markovic}},\ }\bibfield  {title} {\bibinfo {title} {Energy and fuels from
  electrochemical interfaces},\ }\href {https://doi.org/10.1038/nmat4738}
  {\bibfield  {journal} {\bibinfo  {journal} {Nature Materials}\ }\textbf
  {\bibinfo {volume} {16}},\ \bibinfo {pages} {57} (\bibinfo {year}
  {2017})}\BibitemShut {NoStop}%
\bibitem [{\citenamefont {Sundararaman}\ \emph {et~al.}(2022)\citenamefont
  {Sundararaman}, \citenamefont {Vigil-Fowler},\ and\ \citenamefont
  {Schwarz}}]{sundararaman2022improving}%
  \BibitemOpen
  \bibfield  {author} {\bibinfo {author} {\bibfnamefont {R.}~\bibnamefont
  {Sundararaman}}, \bibinfo {author} {\bibfnamefont {D.}~\bibnamefont
  {Vigil-Fowler}},\ and\ \bibinfo {author} {\bibfnamefont {K.}~\bibnamefont
  {Schwarz}},\ }\bibfield  {title} {\bibinfo {title} {Improving the accuracy of
  atomistic simulations of the electrochemical interface},\ }\href@noop {}
  {\bibfield  {journal} {\bibinfo  {journal} {Chemical reviews}\ }\textbf
  {\bibinfo {volume} {122}},\ \bibinfo {pages} {10651} (\bibinfo {year}
  {2022})}\BibitemShut {NoStop}%
\bibitem [{\citenamefont {Deng}\ \emph {et~al.}(2016)\citenamefont {Deng},
  \citenamefont {Novoselov}, \citenamefont {Fu}, \citenamefont {Zheng},
  \citenamefont {Tian},\ and\ \citenamefont {Bao}}]{deng_catalysis_2016}%
  \BibitemOpen
  \bibfield  {author} {\bibinfo {author} {\bibfnamefont {D.}~\bibnamefont
  {Deng}}, \bibinfo {author} {\bibfnamefont {K.~S.}\ \bibnamefont {Novoselov}},
  \bibinfo {author} {\bibfnamefont {Q.}~\bibnamefont {Fu}}, \bibinfo {author}
  {\bibfnamefont {N.}~\bibnamefont {Zheng}}, \bibinfo {author} {\bibfnamefont
  {Z.}~\bibnamefont {Tian}},\ and\ \bibinfo {author} {\bibfnamefont
  {X.}~\bibnamefont {Bao}},\ }\bibfield  {title} {\bibinfo {title} {Catalysis
  with two-dimensional materials and their heterostructures},\ }\href
  {https://doi.org/10.1038/nnano.2015.340} {\bibfield  {journal} {\bibinfo
  {journal} {Nature Nanotechnology}\ }\textbf {\bibinfo {volume} {11}},\
  \bibinfo {pages} {218} (\bibinfo {year} {2016})}\BibitemShut {NoStop}%
\bibitem [{\citenamefont {Li}\ \emph {et~al.}(2023)\citenamefont {Li},
  \citenamefont {Hua}, \citenamefont {Sun}, \citenamefont {Liu}, \citenamefont
  {Li}, \citenamefont {Wang}, \citenamefont {Yang}, \citenamefont {Zhuang},\
  and\ \citenamefont {Wang}}]{li_moire_2023}%
  \BibitemOpen
  \bibfield  {author} {\bibinfo {author} {\bibfnamefont {Y.}~\bibnamefont
  {Li}}, \bibinfo {author} {\bibfnamefont {Y.}~\bibnamefont {Hua}}, \bibinfo
  {author} {\bibfnamefont {N.}~\bibnamefont {Sun}}, \bibinfo {author}
  {\bibfnamefont {S.}~\bibnamefont {Liu}}, \bibinfo {author} {\bibfnamefont
  {H.}~\bibnamefont {Li}}, \bibinfo {author} {\bibfnamefont {C.}~\bibnamefont
  {Wang}}, \bibinfo {author} {\bibfnamefont {X.}~\bibnamefont {Yang}}, \bibinfo
  {author} {\bibfnamefont {Z.}~\bibnamefont {Zhuang}},\ and\ \bibinfo {author}
  {\bibfnamefont {L.}~\bibnamefont {Wang}},\ }\bibfield  {title} {\bibinfo
  {title} {Moiré superlattice engineering of two-dimensional materials for
  electrocatalytic hydrogen evolution reaction},\ }\href
  {https://doi.org/10.1007/s12274-023-5716-9} {\bibfield  {journal} {\bibinfo
  {journal} {Nano Research}\ }\textbf {\bibinfo {volume} {16}},\ \bibinfo
  {pages} {8712} (\bibinfo {year} {2023})}\BibitemShut {NoStop}%
\bibitem [{\citenamefont {Yu}\ \emph {et~al.}(2022)\citenamefont {Yu},
  \citenamefont {Zhang}, \citenamefont {Parks}, \citenamefont {Babar},
  \citenamefont {Carr}, \citenamefont {Craig}, \citenamefont {Van~Winkle},
  \citenamefont {Lyssenko}, \citenamefont {Taniguchi}, \citenamefont
  {Watanabe}, \citenamefont {Viswanathan},\ and\ \citenamefont
  {Bediako}}]{yu_tunable_2022}%
  \BibitemOpen
  \bibfield  {author} {\bibinfo {author} {\bibfnamefont {Y.}~\bibnamefont
  {Yu}}, \bibinfo {author} {\bibfnamefont {K.}~\bibnamefont {Zhang}}, \bibinfo
  {author} {\bibfnamefont {H.}~\bibnamefont {Parks}}, \bibinfo {author}
  {\bibfnamefont {M.}~\bibnamefont {Babar}}, \bibinfo {author} {\bibfnamefont
  {S.}~\bibnamefont {Carr}}, \bibinfo {author} {\bibfnamefont {I.~M.}\
  \bibnamefont {Craig}}, \bibinfo {author} {\bibfnamefont {M.}~\bibnamefont
  {Van~Winkle}}, \bibinfo {author} {\bibfnamefont {A.}~\bibnamefont
  {Lyssenko}}, \bibinfo {author} {\bibfnamefont {T.}~\bibnamefont {Taniguchi}},
  \bibinfo {author} {\bibfnamefont {K.}~\bibnamefont {Watanabe}}, \bibinfo
  {author} {\bibfnamefont {V.}~\bibnamefont {Viswanathan}},\ and\ \bibinfo
  {author} {\bibfnamefont {D.~K.}\ \bibnamefont {Bediako}},\ }\bibfield
  {title} {\bibinfo {title} {Tunable angle-dependent electrochemistry at
  twisted bilayer graphene with moiré flat bands},\ }\href
  {https://doi.org/10.1038/s41557-021-00865-1} {\bibfield  {journal} {\bibinfo
  {journal} {Nature Chemistry}\ }\textbf {\bibinfo {volume} {14}},\ \bibinfo
  {pages} {267} (\bibinfo {year} {2022})}\BibitemShut {NoStop}%
\bibitem [{\citenamefont {Bistritzer}\ and\ \citenamefont
  {MacDonald}(2011)}]{bistritzer_moire_2011}%
  \BibitemOpen
  \bibfield  {author} {\bibinfo {author} {\bibfnamefont {R.}~\bibnamefont
  {Bistritzer}}\ and\ \bibinfo {author} {\bibfnamefont {A.~H.}\ \bibnamefont
  {MacDonald}},\ }\bibfield  {title} {\bibinfo {title} {Moiré bands in twisted
  double-layer graphene},\ }\href {https://doi.org/10.1073/PNAS.1108174108}
  {\bibfield  {journal} {\bibinfo  {journal} {Proceedings of the National
  Academy of Sciences of the United States of America}\ }\textbf {\bibinfo
  {volume} {108}},\ \bibinfo {pages} {12233} (\bibinfo {year}
  {2011})}\BibitemShut {NoStop}%
\bibitem [{\citenamefont {Marcus}(1956)}]{marcus_theory_1956}%
  \BibitemOpen
  \bibfield  {author} {\bibinfo {author} {\bibfnamefont {R.~A.}\ \bibnamefont
  {Marcus}},\ }\bibfield  {title} {\bibinfo {title} {On the theory of
  oxidation-reduction reactions involving electron transfer. {I}},\ }\href
  {https://doi.org/10.1063/1.1742723} {\bibfield  {journal} {\bibinfo
  {journal} {Journal of Chemical Physics}\ }\textbf {\bibinfo {volume} {24}},\
  \bibinfo {pages} {966} (\bibinfo {year} {1956})}\BibitemShut {NoStop}%
\bibitem [{\citenamefont {Marcus}(1957)}]{marcus1957theory}%
  \BibitemOpen
  \bibfield  {author} {\bibinfo {author} {\bibfnamefont {R.}~\bibnamefont
  {Marcus}},\ }\bibfield  {title} {\bibinfo {title} {On the theory of
  oxidation-reduction reactions involving electron transfer. ii. applications
  to data on the rates of isotopic exchange reactions},\ }\href@noop {}
  {\bibfield  {journal} {\bibinfo  {journal} {Journal of Chemical Physics}\
  }\textbf {\bibinfo {volume} {26}},\ \bibinfo {pages} {867} (\bibinfo {year}
  {1957})}\BibitemShut {NoStop}%
\bibitem [{\citenamefont {Marcus}(1965)}]{marcus_theory_1965-1}%
  \BibitemOpen
  \bibfield  {author} {\bibinfo {author} {\bibfnamefont {R.~A.}\ \bibnamefont
  {Marcus}},\ }\bibfield  {title} {\bibinfo {title} {On the {Theory} of
  {Electron}-{Transfer} {Reactions}. {VI}. {Unified} {Treatment} for
  {Homogeneous} and {Electrode} {Reactions}},\ }\href
  {https://doi.org/10.1063/1.1696792} {\bibfield  {journal} {\bibinfo
  {journal} {Journal of Chemical Physics}\ }\textbf {\bibinfo {volume} {43}},\
  \bibinfo {pages} {679} (\bibinfo {year} {1965})}\BibitemShut {NoStop}%
\bibitem [{\citenamefont {Chandler}()}]{chandler_chapter_nodate}%
  \BibitemOpen
  \bibfield  {author} {\bibinfo {author} {\bibfnamefont {D.}~\bibnamefont
  {Chandler}},\ }\href@noop {} {\emph {\bibinfo {title} {{CHAPTER} 2 {Electron}
  transfer in water and other polar environ-ments, how it happens}}},\ \bibinfo
  {type} {Tech. Rep.}\BibitemShut {Stop}%
\bibitem [{\citenamefont {Nitzan}(2014)}]{nitzan_chemical_2014}%
  \BibitemOpen
  \bibfield  {author} {\bibinfo {author} {\bibfnamefont {A.}~\bibnamefont
  {Nitzan}},\ }\href@noop {} {\emph {\bibinfo {title} {Chemical {Dynamics} in
  {Condensed} {Phases}: {Relaxation}, {Transfer}, and {Reactions} in
  {Condensed} {Molecular} {Systems}}}},\ Oxford {Graduate} {Texts}\ (\bibinfo
  {publisher} {Oxford University Press},\ \bibinfo {address} {Oxford, New
  York},\ \bibinfo {year} {2014})\BibitemShut {NoStop}%
\bibitem [{\citenamefont {Ferrario}\ \emph {et~al.}(2006)\citenamefont
  {Ferrario}, \citenamefont {Ciccotti},\ and\ \citenamefont
  {Binder}}]{ferrario_computer_2006}%
  \BibitemOpen
  \bibinfo {editor} {\bibfnamefont {M.}~\bibnamefont {Ferrario}}, \bibinfo
  {editor} {\bibfnamefont {G.}~\bibnamefont {Ciccotti}},\ and\ \bibinfo
  {editor} {\bibfnamefont {K.}~\bibnamefont {Binder}},\ eds.,\ \href
  {https://doi.org/10.1007/3-540-35284-8} {\emph {\bibinfo {title} {Computer
  {Simulations} in {Condensed} {Matter} {Systems}: {From} {Materials} to
  {Chemical} {Biology} {Volume} 2}}},\ Vol.\ \bibinfo {volume} {704}\ (\bibinfo
   {publisher} {Springer Berlin Heidelberg},\ \bibinfo {address} {Berlin,
  Heidelberg},\ \bibinfo {year} {2006})\BibitemShut {NoStop}%
\bibitem [{\citenamefont {Wong}\ \emph {et~al.}(2020)\citenamefont {Wong},
  \citenamefont {Nuckolls}, \citenamefont {Oh}, \citenamefont {Lian},
  \citenamefont {Xie}, \citenamefont {Jeon}, \citenamefont {Watanabe},
  \citenamefont {Taniguchi}, \citenamefont {Bernevig},\ and\ \citenamefont
  {Yazdani}}]{wong_cascade_2020}%
  \BibitemOpen
  \bibfield  {author} {\bibinfo {author} {\bibfnamefont {D.}~\bibnamefont
  {Wong}}, \bibinfo {author} {\bibfnamefont {K.~P.}\ \bibnamefont {Nuckolls}},
  \bibinfo {author} {\bibfnamefont {M.}~\bibnamefont {Oh}}, \bibinfo {author}
  {\bibfnamefont {B.}~\bibnamefont {Lian}}, \bibinfo {author} {\bibfnamefont
  {Y.}~\bibnamefont {Xie}}, \bibinfo {author} {\bibfnamefont {S.}~\bibnamefont
  {Jeon}}, \bibinfo {author} {\bibfnamefont {K.}~\bibnamefont {Watanabe}},
  \bibinfo {author} {\bibfnamefont {T.}~\bibnamefont {Taniguchi}}, \bibinfo
  {author} {\bibfnamefont {B.~A.}\ \bibnamefont {Bernevig}},\ and\ \bibinfo
  {author} {\bibfnamefont {A.}~\bibnamefont {Yazdani}},\ }\bibfield  {title}
  {\bibinfo {title} {Cascade of electronic transitions in magic-angle twisted
  bilayer graphene},\ }\href {https://doi.org/10.1038/s41586-020-2339-0}
  {\bibfield  {journal} {\bibinfo  {journal} {Nature}\ }\textbf {\bibinfo
  {volume} {582}},\ \bibinfo {pages} {198} (\bibinfo {year}
  {2020})}\BibitemShut {NoStop}%
\bibitem [{\citenamefont {Ashcroft}\ and\ \citenamefont
  {Mermin}(1976)}]{ashcroft_solid_1976}%
  \BibitemOpen
  \bibfield  {author} {\bibinfo {author} {\bibfnamefont {N.~W.}\ \bibnamefont
  {Ashcroft}}\ and\ \bibinfo {author} {\bibfnamefont {N.~D.}\ \bibnamefont
  {Mermin}},\ }\href@noop {} {\emph {\bibinfo {title} {Solid state physics}}}\
  (\bibinfo  {publisher} {Brooks/Cole},\ \bibinfo {address} {Australia},\
  \bibinfo {year} {1976})\BibitemShut {NoStop}%
\bibitem [{\citenamefont {Noori}\ \emph {et~al.}(2019)\citenamefont {Noori},
  \citenamefont {Cheng}, \citenamefont {Xuan},\ and\ \citenamefont
  {Quek}}]{noori_dielectric_2019}%
  \BibitemOpen
  \bibfield  {author} {\bibinfo {author} {\bibfnamefont {K.}~\bibnamefont
  {Noori}}, \bibinfo {author} {\bibfnamefont {N.~L.~Q.}\ \bibnamefont {Cheng}},
  \bibinfo {author} {\bibfnamefont {F.}~\bibnamefont {Xuan}},\ and\ \bibinfo
  {author} {\bibfnamefont {S.~Y.}\ \bibnamefont {Quek}},\ }\bibfield  {title}
  {\bibinfo {title} {Dielectric screening by {2D} substrates},\ }\href
  {https://doi.org/10.1088/2053-1583/ab1e06} {\bibfield  {journal} {\bibinfo
  {journal} {2D Materials}\ }\textbf {\bibinfo {volume} {6}},\ \bibinfo {pages}
  {035036} (\bibinfo {year} {2019})}\BibitemShut {NoStop}%
\bibitem [{\citenamefont {Ando}\ \emph {et~al.}(1982)\citenamefont {Ando},
  \citenamefont {Fowler},\ and\ \citenamefont {Stern}}]{ando_electronic_1982}%
  \BibitemOpen
  \bibfield  {author} {\bibinfo {author} {\bibfnamefont {T.}~\bibnamefont
  {Ando}}, \bibinfo {author} {\bibfnamefont {A.~B.}\ \bibnamefont {Fowler}},\
  and\ \bibinfo {author} {\bibfnamefont {F.}~\bibnamefont {Stern}},\ }\bibfield
   {title} {\bibinfo {title} {Electronic properties of two-dimensional
  systems},\ }\href {https://doi.org/10.1103/RevModPhys.54.437} {\bibfield
  {journal} {\bibinfo  {journal} {Reviews of Modern Physics}\ }\textbf
  {\bibinfo {volume} {54}},\ \bibinfo {pages} {437} (\bibinfo {year}
  {1982})}\BibitemShut {NoStop}%
\bibitem [{SI()}]{SI}%
  \BibitemOpen
  \href@noop {} {}\bibinfo {note} {Supporting information}\BibitemShut
  {NoStop}%
\bibitem [{\citenamefont {Warshel}(1982)}]{warshel_dynamics_1982}%
  \BibitemOpen
  \bibfield  {author} {\bibinfo {author} {\bibfnamefont {A.}~\bibnamefont
  {Warshel}},\ }\bibfield  {title} {\bibinfo {title} {Dynamics of reactions in
  polar solvents. {Semiclassical} trajectory studies of electron-transfer and
  proton-transfer reactions},\ }\href {https://doi.org/10.1021/j100209a016}
  {\bibfield  {journal} {\bibinfo  {journal} {The Journal of Physical
  Chemistry}\ }\textbf {\bibinfo {volume} {86}},\ \bibinfo {pages} {2218}
  (\bibinfo {year} {1982})}\BibitemShut {NoStop}%
\bibitem [{\citenamefont {Reed}\ \emph {et~al.}(2007)\citenamefont {Reed},
  \citenamefont {Lanning},\ and\ \citenamefont
  {Madden}}]{reed_electrochemical_2007}%
  \BibitemOpen
  \bibfield  {author} {\bibinfo {author} {\bibfnamefont {S.~K.}\ \bibnamefont
  {Reed}}, \bibinfo {author} {\bibfnamefont {O.~J.}\ \bibnamefont {Lanning}},\
  and\ \bibinfo {author} {\bibfnamefont {P.~A.}\ \bibnamefont {Madden}},\
  }\bibfield  {title} {\bibinfo {title} {Electrochemical interface between an
  ionic liquid and a model metallic electrode},\ }\href
  {https://doi.org/10.1063/1.2464084} {\bibfield  {journal} {\bibinfo
  {journal} {Journal of Chemical Physics}\ }\textbf {\bibinfo {volume} {126}},\
  \bibinfo {pages} {084704} (\bibinfo {year} {2007})}\BibitemShut {NoStop}%
\bibitem [{\citenamefont {Reed}\ \emph {et~al.}(2008)\citenamefont {Reed},
  \citenamefont {Madden},\ and\ \citenamefont
  {Papadopoulos}}]{reed_electrochemical_2008}%
  \BibitemOpen
  \bibfield  {author} {\bibinfo {author} {\bibfnamefont {S.~K.}\ \bibnamefont
  {Reed}}, \bibinfo {author} {\bibfnamefont {P.~A.}\ \bibnamefont {Madden}},\
  and\ \bibinfo {author} {\bibfnamefont {A.}~\bibnamefont {Papadopoulos}},\
  }\bibfield  {title} {\bibinfo {title} {Electrochemical charge transfer at a
  metallic electrode: {A} simulation study},\ }\href
  {https://doi.org/10.1063/1.2844801} {\bibfield  {journal} {\bibinfo
  {journal} {Journal of Chemical Physics}\ }\textbf {\bibinfo {volume} {128}},\
  \bibinfo {pages} {124701} (\bibinfo {year} {2008})}\BibitemShut {NoStop}%
\bibitem [{\citenamefont {Willard}\ \emph {et~al.}(2008)\citenamefont
  {Willard}, \citenamefont {Reed}, \citenamefont {Madden},\ and\ \citenamefont
  {Chandler}}]{willard_water_2008}%
  \BibitemOpen
  \bibfield  {author} {\bibinfo {author} {\bibfnamefont {A.~P.}\ \bibnamefont
  {Willard}}, \bibinfo {author} {\bibfnamefont {S.~K.}\ \bibnamefont {Reed}},
  \bibinfo {author} {\bibfnamefont {P.~A.}\ \bibnamefont {Madden}},\ and\
  \bibinfo {author} {\bibfnamefont {D.}~\bibnamefont {Chandler}},\ }\bibfield
  {title} {\bibinfo {title} {Water at an electrochemical interface—a
  simulation study},\ }\href {https://doi.org/10.1039/B805544K} {\bibfield
  {journal} {\bibinfo  {journal} {Faraday Discussions}\ }\textbf {\bibinfo
  {volume} {141}},\ \bibinfo {pages} {423} (\bibinfo {year}
  {2008})}\BibitemShut {NoStop}%
\bibitem [{\citenamefont {Siepmann}\ and\ \citenamefont
  {Sprik}(1995)}]{siepmann1995influence}%
  \BibitemOpen
  \bibfield  {author} {\bibinfo {author} {\bibfnamefont {J.~I.}\ \bibnamefont
  {Siepmann}}\ and\ \bibinfo {author} {\bibfnamefont {M.}~\bibnamefont
  {Sprik}},\ }\bibfield  {title} {\bibinfo {title} {Influence of surface
  topology and electrostatic potential on water/electrode systems},\
  }\href@noop {} {\bibfield  {journal} {\bibinfo  {journal} {Journal of
  chemical physics}\ }\textbf {\bibinfo {volume} {102}},\ \bibinfo {pages}
  {511} (\bibinfo {year} {1995})}\BibitemShut {NoStop}%
\bibitem [{\citenamefont {Scalfi}\ \emph
  {et~al.}(2020{\natexlab{a}})\citenamefont {Scalfi}, \citenamefont {Limmer},
  \citenamefont {Alessandro~Coretti}, \citenamefont {Sara~Bonella},
  \citenamefont {Madden}, \citenamefont {Salanne~ai},\ and\ \citenamefont
  {Rotenberg}}]{scalfi_charge_2020}%
  \BibitemOpen
  \bibfield  {author} {\bibinfo {author} {\bibfnamefont {L.}~\bibnamefont
  {Scalfi}}, \bibinfo {author} {\bibfnamefont {D.~T.}\ \bibnamefont {Limmer}},
  \bibinfo {author} {\bibfnamefont {b.}~\bibnamefont {Alessandro~Coretti}},
  \bibinfo {author} {\bibfnamefont {f.}~\bibnamefont {Sara~Bonella}}, \bibinfo
  {author} {\bibfnamefont {P.~A.}\ \bibnamefont {Madden}}, \bibinfo {author}
  {\bibfnamefont {M.}~\bibnamefont {Salanne~ai}},\ and\ \bibinfo {author}
  {\bibfnamefont {B.}~\bibnamefont {Rotenberg}},\ }\bibfield  {title} {\bibinfo
  {title} {Charge fluctuations from molecular simulations in the
  constant-potential ensemble},\ }\href {https://doi.org/10.1039/c9cp06285h}
  {\bibfield  {journal} {\bibinfo  {journal} {Phys. Chem. Chem. Phys}\ }\textbf
  {\bibinfo {volume} {22}},\ \bibinfo {pages} {10480} (\bibinfo {year}
  {2020}{\natexlab{a}})}\BibitemShut {NoStop}%
\bibitem [{\citenamefont {Limmer}\ \emph {et~al.}(2013)\citenamefont {Limmer},
  \citenamefont {Merlet}, \citenamefont {Salanne}, \citenamefont {Chandler},
  \citenamefont {Madden}, \citenamefont {Van~Roij},\ and\ \citenamefont
  {Rotenberg}}]{limmer2013charge}%
  \BibitemOpen
  \bibfield  {author} {\bibinfo {author} {\bibfnamefont {D.~T.}\ \bibnamefont
  {Limmer}}, \bibinfo {author} {\bibfnamefont {C.}~\bibnamefont {Merlet}},
  \bibinfo {author} {\bibfnamefont {M.}~\bibnamefont {Salanne}}, \bibinfo
  {author} {\bibfnamefont {D.}~\bibnamefont {Chandler}}, \bibinfo {author}
  {\bibfnamefont {P.~A.}\ \bibnamefont {Madden}}, \bibinfo {author}
  {\bibfnamefont {R.}~\bibnamefont {Van~Roij}},\ and\ \bibinfo {author}
  {\bibfnamefont {B.}~\bibnamefont {Rotenberg}},\ }\bibfield  {title} {\bibinfo
  {title} {Charge fluctuations in nanoscale capacitors},\ }\href@noop {}
  {\bibfield  {journal} {\bibinfo  {journal} {Physical review letters}\
  }\textbf {\bibinfo {volume} {111}},\ \bibinfo {pages} {106102} (\bibinfo
  {year} {2013})}\BibitemShut {NoStop}%
\bibitem [{\citenamefont {Scalfi}\ \emph {et~al.}(2021)\citenamefont {Scalfi},
  \citenamefont {Salanne},\ and\ \citenamefont
  {Rotenberg}}]{scalfi_molecular_2021}%
  \BibitemOpen
  \bibfield  {author} {\bibinfo {author} {\bibfnamefont {L.}~\bibnamefont
  {Scalfi}}, \bibinfo {author} {\bibfnamefont {M.}~\bibnamefont {Salanne}},\
  and\ \bibinfo {author} {\bibfnamefont {B.}~\bibnamefont {Rotenberg}},\
  }\bibfield  {title} {\bibinfo {title} {Molecular simulation of
  electrode-solution interfaces},\ }\href@noop {} {\bibfield  {journal}
  {\bibinfo  {journal} {Annual Review of Physical Chemistry}\ }\textbf
  {\bibinfo {volume} {72}},\ \bibinfo {pages} {189} (\bibinfo {year}
  {2021})}\BibitemShut {NoStop}%
\bibitem [{\citenamefont {Scalfi}\ \emph
  {et~al.}(2020{\natexlab{b}})\citenamefont {Scalfi}, \citenamefont {Dufils},
  \citenamefont {Reeves}, \citenamefont {Rotenberg},\ and\ \citenamefont
  {Salanne}}]{scalfi_semiclassical_2020}%
  \BibitemOpen
  \bibfield  {author} {\bibinfo {author} {\bibfnamefont {L.}~\bibnamefont
  {Scalfi}}, \bibinfo {author} {\bibfnamefont {T.}~\bibnamefont {Dufils}},
  \bibinfo {author} {\bibfnamefont {K.~G.}\ \bibnamefont {Reeves}}, \bibinfo
  {author} {\bibfnamefont {B.}~\bibnamefont {Rotenberg}},\ and\ \bibinfo
  {author} {\bibfnamefont {M.}~\bibnamefont {Salanne}},\ }\bibfield  {title}
  {\bibinfo {title} {A semiclassical {Thomas}–{Fermi} model to tune the
  metallicity of electrodes in molecular simulations},\ }\href
  {https://doi.org/10.1063/5.0028232} {\bibfield  {journal} {\bibinfo
  {journal} {Journal of Chemical Physics}\ }\textbf {\bibinfo {volume} {153}},\
  \bibinfo {pages} {174704} (\bibinfo {year} {2020}{\natexlab{b}})}\BibitemShut
  {NoStop}%
\bibitem [{\citenamefont {Scalfi}\ and\ \citenamefont
  {Rotenberg}(2021)}]{scalfi_microscopic_2021}%
  \BibitemOpen
  \bibfield  {author} {\bibinfo {author} {\bibfnamefont {L.}~\bibnamefont
  {Scalfi}}\ and\ \bibinfo {author} {\bibfnamefont {B.}~\bibnamefont
  {Rotenberg}},\ }\bibfield  {title} {\bibinfo {title} {Microscopic origin of
  the effect of substrate metallicity on interfacial free energies},\ }\href
  {https://doi.org/10.1073/pnas.2108769118} {\bibfield  {journal} {\bibinfo
  {journal} {Proceedings of the National Academy of Sciences}\ }\textbf
  {\bibinfo {volume} {118}},\ \bibinfo {pages} {e2108769118} (\bibinfo {year}
  {2021})}\BibitemShut {NoStop}%
\bibitem [{\citenamefont {Hwang}\ and\ \citenamefont
  {Warshel}(1987)}]{hwang_microscopic_1987}%
  \BibitemOpen
  \bibfield  {author} {\bibinfo {author} {\bibfnamefont {J.~K.}\ \bibnamefont
  {Hwang}}\ and\ \bibinfo {author} {\bibfnamefont {A.}~\bibnamefont
  {Warshel}},\ }\bibfield  {title} {\bibinfo {title} {Microscopic examination
  of free-energy relationships for electron transfer in polar solvents},\
  }\href {https://doi.org/10.1021/ja00237a013} {\bibfield  {journal} {\bibinfo
  {journal} {Journal of the American Chemical Society}\ }\textbf {\bibinfo
  {volume} {109}},\ \bibinfo {pages} {715} (\bibinfo {year}
  {1987})}\BibitemShut {NoStop}%
\bibitem [{\citenamefont {King}\ and\ \citenamefont
  {Warshel}(1990)}]{king_investigation_1990}%
  \BibitemOpen
  \bibfield  {author} {\bibinfo {author} {\bibfnamefont {G.}~\bibnamefont
  {King}}\ and\ \bibinfo {author} {\bibfnamefont {A.}~\bibnamefont {Warshel}},\
  }\bibfield  {title} {\bibinfo {title} {Investigation of the free energy
  functions for electron transfer reactions},\ }\href
  {https://doi.org/10.1063/1.459255} {\bibfield  {journal} {\bibinfo  {journal}
  {Journal of Chemical Physics}\ }\textbf {\bibinfo {volume} {93}},\ \bibinfo
  {pages} {8682} (\bibinfo {year} {1990})}\BibitemShut {NoStop}%
\bibitem [{\citenamefont {Shirts}\ and\ \citenamefont
  {Chodera}(2008)}]{shirts2008statistically}%
  \BibitemOpen
  \bibfield  {author} {\bibinfo {author} {\bibfnamefont {M.~R.}\ \bibnamefont
  {Shirts}}\ and\ \bibinfo {author} {\bibfnamefont {J.~D.}\ \bibnamefont
  {Chodera}},\ }\bibfield  {title} {\bibinfo {title} {Statistically optimal
  analysis of samples from multiple equilibrium states},\ }\href@noop {}
  {\bibfield  {journal} {\bibinfo  {journal} {Journal of chemical physics}\
  }\textbf {\bibinfo {volume} {129}} (\bibinfo {year} {2008})}\BibitemShut
  {NoStop}%
\bibitem [{\citenamefont {Thompson}\ \emph {et~al.}(2022)\citenamefont
  {Thompson}, \citenamefont {Aktulga}, \citenamefont {Berger}, \citenamefont
  {Bolintineanu}, \citenamefont {Brown}, \citenamefont {Crozier}, \citenamefont
  {in~'t Veld}, \citenamefont {Kohlmeyer}, \citenamefont {Moore}, \citenamefont
  {Nguyen}, \citenamefont {Shan}, \citenamefont {Stevens}, \citenamefont
  {Tranchida}, \citenamefont {Trott},\ and\ \citenamefont
  {Plimpton}}]{thompson_lammps_2022}%
  \BibitemOpen
  \bibfield  {author} {\bibinfo {author} {\bibfnamefont {A.~P.}\ \bibnamefont
  {Thompson}}, \bibinfo {author} {\bibfnamefont {H.~M.}\ \bibnamefont
  {Aktulga}}, \bibinfo {author} {\bibfnamefont {R.}~\bibnamefont {Berger}},
  \bibinfo {author} {\bibfnamefont {D.~S.}\ \bibnamefont {Bolintineanu}},
  \bibinfo {author} {\bibfnamefont {W.~M.}\ \bibnamefont {Brown}}, \bibinfo
  {author} {\bibfnamefont {P.~S.}\ \bibnamefont {Crozier}}, \bibinfo {author}
  {\bibfnamefont {P.~J.}\ \bibnamefont {in~'t Veld}}, \bibinfo {author}
  {\bibfnamefont {A.}~\bibnamefont {Kohlmeyer}}, \bibinfo {author}
  {\bibfnamefont {S.~G.}\ \bibnamefont {Moore}}, \bibinfo {author}
  {\bibfnamefont {T.~D.}\ \bibnamefont {Nguyen}}, \bibinfo {author}
  {\bibfnamefont {R.}~\bibnamefont {Shan}}, \bibinfo {author} {\bibfnamefont
  {M.~J.}\ \bibnamefont {Stevens}}, \bibinfo {author} {\bibfnamefont
  {J.}~\bibnamefont {Tranchida}}, \bibinfo {author} {\bibfnamefont
  {C.}~\bibnamefont {Trott}},\ and\ \bibinfo {author} {\bibfnamefont {S.~J.}\
  \bibnamefont {Plimpton}},\ }\bibfield  {title} {\bibinfo {title} {{LAMMPS} -
  a flexible simulation tool for particle-based materials modeling at the
  atomic, meso, and continuum scales},\ }\href
  {https://doi.org/10.1016/j.cpc.2021.108171} {\bibfield  {journal} {\bibinfo
  {journal} {Computer Physics Communications}\ }\textbf {\bibinfo {volume}
  {271}},\ \bibinfo {pages} {108171} (\bibinfo {year} {2022})}\BibitemShut
  {NoStop}%
\bibitem [{\citenamefont {Ahrens-Iwers}\ \emph {et~al.}(2022)\citenamefont
  {Ahrens-Iwers}, \citenamefont {Janssen}, \citenamefont {Tee},\ and\
  \citenamefont {Meißner}}]{ahrens-iwers_electrode_2022}%
  \BibitemOpen
  \bibfield  {author} {\bibinfo {author} {\bibfnamefont {L.~J.}\ \bibnamefont
  {Ahrens-Iwers}}, \bibinfo {author} {\bibfnamefont {M.}~\bibnamefont
  {Janssen}}, \bibinfo {author} {\bibfnamefont {S.~R.}\ \bibnamefont {Tee}},\
  and\ \bibinfo {author} {\bibfnamefont {R.~H.}\ \bibnamefont {Meißner}},\
  }\bibfield  {title} {\bibinfo {title} {{ELECTRODE}: {An} electrochemistry
  package for atomistic simulations},\ }\href
  {https://doi.org/10.1063/5.0099239/2841985} {\bibfield  {journal} {\bibinfo
  {journal} {Journal of Chemical Physics}\ }\textbf {\bibinfo {volume} {157}},\
  \bibinfo {pages} {84801} (\bibinfo {year} {2022})}\BibitemShut {NoStop}%
\bibitem [{\citenamefont {Berendsen}\ \emph {et~al.}(1987)\citenamefont
  {Berendsen}, \citenamefont {Grigera},\ and\ \citenamefont
  {Straatsma}}]{berendsen_missing_1987}%
  \BibitemOpen
  \bibfield  {author} {\bibinfo {author} {\bibfnamefont {H.~J.~C.}\
  \bibnamefont {Berendsen}}, \bibinfo {author} {\bibfnamefont {J.~R.}\
  \bibnamefont {Grigera}},\ and\ \bibinfo {author} {\bibfnamefont {T.~P.}\
  \bibnamefont {Straatsma}},\ }\bibfield  {title} {\bibinfo {title} {The
  missing term in effective pair potentials},\ }\href
  {https://doi.org/10.1021/j100308a038} {\bibfield  {journal} {\bibinfo
  {journal} {The Journal of Physical Chemistry}\ }\textbf {\bibinfo {volume}
  {91}},\ \bibinfo {pages} {6269} (\bibinfo {year} {1987})}\BibitemShut
  {NoStop}%
\bibitem [{\citenamefont {Rami~Reddy}\ and\ \citenamefont
  {Berkowitz}(1989)}]{rami_reddy_dielectric_1989}%
  \BibitemOpen
  \bibfield  {author} {\bibinfo {author} {\bibfnamefont {M.}~\bibnamefont
  {Rami~Reddy}}\ and\ \bibinfo {author} {\bibfnamefont {M.}~\bibnamefont
  {Berkowitz}},\ }\bibfield  {title} {\bibinfo {title} {The dielectric constant
  of {SPC}/{E} water},\ }\href {https://doi.org/10.1016/0009-2614(89)85344-8}
  {\bibfield  {journal} {\bibinfo  {journal} {Chemical Physics Letters}\
  }\textbf {\bibinfo {volume} {155}},\ \bibinfo {pages} {173} (\bibinfo {year}
  {1989})}\BibitemShut {NoStop}%
\bibitem [{\citenamefont {Dufils}\ \emph {et~al.}(2019)\citenamefont {Dufils},
  \citenamefont {Jeanmairet}, \citenamefont {Rotenberg}, \citenamefont
  {Sprik},\ and\ \citenamefont {Salanne}}]{dufils_simulating_2019}%
  \BibitemOpen
  \bibfield  {author} {\bibinfo {author} {\bibfnamefont {T.}~\bibnamefont
  {Dufils}}, \bibinfo {author} {\bibfnamefont {G.}~\bibnamefont {Jeanmairet}},
  \bibinfo {author} {\bibfnamefont {B.}~\bibnamefont {Rotenberg}}, \bibinfo
  {author} {\bibfnamefont {M.}~\bibnamefont {Sprik}},\ and\ \bibinfo {author}
  {\bibfnamefont {M.}~\bibnamefont {Salanne}},\ }\bibfield  {title} {\bibinfo
  {title} {Simulating {Electrochemical} {Systems} by {Combining} the {Finite}
  {Field} {Method} with a {Constant} {Potential} {Electrode}},\ }\href
  {https://doi.org/10.1103/PHYSREVLETT.123.195501/FIGURES/3/MEDIUM} {\bibfield
  {journal} {\bibinfo  {journal} {Physical Review Letters}\ }\textbf {\bibinfo
  {volume} {123}},\ \bibinfo {pages} {195501} (\bibinfo {year}
  {2019})}\BibitemShut {NoStop}%
\bibitem [{\citenamefont {Li}\ \emph {et~al.}(2013)\citenamefont {Li},
  \citenamefont {Roberts}, \citenamefont {Chakravorty},\ and\ \citenamefont
  {Merz}}]{li_rational_2013}%
  \BibitemOpen
  \bibfield  {author} {\bibinfo {author} {\bibfnamefont {P.}~\bibnamefont
  {Li}}, \bibinfo {author} {\bibfnamefont {B.~P.}\ \bibnamefont {Roberts}},
  \bibinfo {author} {\bibfnamefont {D.~K.}\ \bibnamefont {Chakravorty}},\ and\
  \bibinfo {author} {\bibfnamefont {K.~M.~J.}\ \bibnamefont {Merz}},\
  }\bibfield  {title} {\bibinfo {title} {Rational {Design} of {Particle} {Mesh}
  {Ewald} {Compatible} {Lennard}-{Jones} {Parameters} for +2 {Metal} {Cations}
  in {Explicit} {Solvent}},\ }\href {https://doi.org/10.1021/ct400146w}
  {\bibfield  {journal} {\bibinfo  {journal} {Journal of Chemical Theory and
  Computation}\ }\textbf {\bibinfo {volume} {9}},\ \bibinfo {pages} {2733}
  (\bibinfo {year} {2013})}\BibitemShut {NoStop}%
\bibitem [{\citenamefont {Hummer}\ \emph {et~al.}(2001)\citenamefont {Hummer},
  \citenamefont {Rasaiah},\ and\ \citenamefont {Noworyta}}]{hummer_water_2001}%
  \BibitemOpen
  \bibfield  {author} {\bibinfo {author} {\bibfnamefont {G.}~\bibnamefont
  {Hummer}}, \bibinfo {author} {\bibfnamefont {J.~C.}\ \bibnamefont
  {Rasaiah}},\ and\ \bibinfo {author} {\bibfnamefont {J.~P.}\ \bibnamefont
  {Noworyta}},\ }\bibfield  {title} {\bibinfo {title} {Water conduction through
  the hydrophobic channel of a carbon nanotube},\ }\href
  {https://doi.org/10.1038/35102535} {\bibfield  {journal} {\bibinfo  {journal}
  {Nature}\ }\textbf {\bibinfo {volume} {414}},\ \bibinfo {pages} {188}
  (\bibinfo {year} {2001})}\BibitemShut {NoStop}%
\bibitem [{\citenamefont {Yagasaki}\ \emph {et~al.}(2020)\citenamefont
  {Yagasaki}, \citenamefont {Matsumoto},\ and\ \citenamefont
  {Tanaka}}]{yagasaki_lennard-jones_2020}%
  \BibitemOpen
  \bibfield  {author} {\bibinfo {author} {\bibfnamefont {T.}~\bibnamefont
  {Yagasaki}}, \bibinfo {author} {\bibfnamefont {M.}~\bibnamefont
  {Matsumoto}},\ and\ \bibinfo {author} {\bibfnamefont {H.}~\bibnamefont
  {Tanaka}},\ }\bibfield  {title} {\bibinfo {title} {Lennard-{Jones}
  {Parameters} {Determined} to {Reproduce} the {Solubility} of {NaCl} and {KCl}
  in {SPC}/{E}, {TIP3P}, and {TIP4P}/2005 {Water}},\ }\href
  {https://doi.org/10.1021/acs.jctc.9b00941} {\bibfield  {journal} {\bibinfo
  {journal} {Journal of Chemical Theory and Computation}\ }\textbf {\bibinfo
  {volume} {16}},\ \bibinfo {pages} {2460} (\bibinfo {year}
  {2020})}\BibitemShut {NoStop}%
\bibitem [{\citenamefont {Curtiss}\ \emph {et~al.}(1986)\citenamefont
  {Curtiss}, \citenamefont {Woods~Halley}, \citenamefont {Hautman},\ and\
  \citenamefont {Rahman}}]{curtiss_nonadditivity_1986}%
  \BibitemOpen
  \bibfield  {author} {\bibinfo {author} {\bibfnamefont {L.~A.}\ \bibnamefont
  {Curtiss}}, \bibinfo {author} {\bibfnamefont {J.}~\bibnamefont
  {Woods~Halley}}, \bibinfo {author} {\bibfnamefont {J.}~\bibnamefont
  {Hautman}},\ and\ \bibinfo {author} {\bibfnamefont {A.}~\bibnamefont
  {Rahman}},\ }\bibfield  {title} {\bibinfo {title} {Nonadditivity of ab initio
  pair potentials for molecular dynamics of multivalent transition metal ions
  in water},\ }\href {https://doi.org/10.1063/1.452130} {\bibfield  {journal}
  {\bibinfo  {journal} {Journal of Chemical Physics}\ }\textbf {\bibinfo
  {volume} {86}},\ \bibinfo {pages} {2319} (\bibinfo {year}
  {1986})}\BibitemShut {NoStop}%
\bibitem [{\citenamefont {Hutchinson}\ \emph {et~al.}(1999)\citenamefont
  {Hutchinson}, \citenamefont {Walters}, \citenamefont {Rowley},\ and\
  \citenamefont {Madden}}]{hutchinson_ionic_1999}%
  \BibitemOpen
  \bibfield  {author} {\bibinfo {author} {\bibfnamefont {F.}~\bibnamefont
  {Hutchinson}}, \bibinfo {author} {\bibfnamefont {M.~K.}\ \bibnamefont
  {Walters}}, \bibinfo {author} {\bibfnamefont {A.~J.}\ \bibnamefont
  {Rowley}},\ and\ \bibinfo {author} {\bibfnamefont {P.~A.}\ \bibnamefont
  {Madden}},\ }\bibfield  {title} {\bibinfo {title} {The “ionic” to
  “molecular” transitions in {AlCl3} and {FeCl3} as predicted by an ionic
  interaction model},\ }\href {https://doi.org/10.1063/1.478480} {\bibfield
  {journal} {\bibinfo  {journal} {Journal of Chemical Physics}\ }\textbf
  {\bibinfo {volume} {110}},\ \bibinfo {pages} {5821} (\bibinfo {year}
  {1999})}\BibitemShut {NoStop}%
\bibitem [{\citenamefont {Smith}\ and\ \citenamefont
  {Halley}(1994)}]{smith_simulation_1994}%
  \BibitemOpen
  \bibfield  {author} {\bibinfo {author} {\bibfnamefont {B.~B.}\ \bibnamefont
  {Smith}}\ and\ \bibinfo {author} {\bibfnamefont {J.~W.}\ \bibnamefont
  {Halley}},\ }\bibfield  {title} {\bibinfo {title} {Simulation study of the
  ferrous ferric electron transfer at a metal-aqueous electrolyte interface},\
  }\href {https://doi.org/10.1063/1.467841} {\bibfield  {journal} {\bibinfo
  {journal} {J. Chem. Phys}\ }\textbf {\bibinfo {volume} {101}},\ \bibinfo
  {pages} {10915} (\bibinfo {year} {1994})}\BibitemShut {NoStop}%
\bibitem [{\citenamefont {Cox}\ \emph {et~al.}(2021)\citenamefont {Cox},
  \citenamefont {Kranthi}, \citenamefont {Mandadapu}, \citenamefont
  {Geissler},\ and\ \citenamefont {Mandadapu}}]{cox_quadrupole-mediated_2021}%
  \BibitemOpen
  \bibfield  {author} {\bibinfo {author} {\bibfnamefont {S.~J.}\ \bibnamefont
  {Cox}}, \bibinfo {author} {\bibfnamefont {.}~\bibnamefont {Kranthi}},
  \bibinfo {author} {\bibfnamefont {K.}~\bibnamefont {Mandadapu}}, \bibinfo
  {author} {\bibfnamefont {P.~L.}\ \bibnamefont {Geissler}},\ and\ \bibinfo
  {author} {\bibfnamefont {K.~K.}\ \bibnamefont {Mandadapu}},\ }\bibfield
  {title} {\bibinfo {title} {Quadrupole-mediated dielectric response and the
  charge-asymmetric solvation of ions in water},\ }\href
  {https://doi.org/10.1063/5.0051399} {\bibfield  {journal} {\bibinfo
  {journal} {J. Chem. Phys}\ }\textbf {\bibinfo {volume} {154}},\ \bibinfo
  {pages} {244502} (\bibinfo {year} {2021})}\BibitemShut {NoStop}%
\bibitem [{\citenamefont {Cox}\ and\ \citenamefont
  {Geissler}(2022)}]{cox2022dielectric}%
  \BibitemOpen
  \bibfield  {author} {\bibinfo {author} {\bibfnamefont {S.~J.}\ \bibnamefont
  {Cox}}\ and\ \bibinfo {author} {\bibfnamefont {P.~L.}\ \bibnamefont
  {Geissler}},\ }\bibfield  {title} {\bibinfo {title} {Dielectric response of
  thin water films: a thermodynamic perspective},\ }\href@noop {} {\bibfield
  {journal} {\bibinfo  {journal} {Chemical Science}\ }\textbf {\bibinfo
  {volume} {13}},\ \bibinfo {pages} {9102} (\bibinfo {year}
  {2022})}\BibitemShut {NoStop}%
\bibitem [{\citenamefont {Nair}\ \emph {et~al.}(2024)\citenamefont {Nair},
  \citenamefont {Pireddu},\ and\ \citenamefont {Rotenberg}}]{nair2024induced}%
  \BibitemOpen
  \bibfield  {author} {\bibinfo {author} {\bibfnamefont {S.}~\bibnamefont
  {Nair}}, \bibinfo {author} {\bibfnamefont {G.}~\bibnamefont {Pireddu}},\ and\
  \bibinfo {author} {\bibfnamefont {B.}~\bibnamefont {Rotenberg}},\ }\bibfield
  {title} {\bibinfo {title} {Induced charges in a thomas--fermi metal: insights
  from molecular simulations},\ }\href@noop {} {\bibfield  {journal} {\bibinfo
  {journal} {Molecular Physics}\ ,\ \bibinfo {pages} {e2365990}} (\bibinfo
  {year} {2024})}\BibitemShut {NoStop}%
\bibitem [{\citenamefont {Liu}\ and\ \citenamefont
  {Newton}(1994)}]{liu1994reorganization}%
  \BibitemOpen
  \bibfield  {author} {\bibinfo {author} {\bibfnamefont {Y.-P.}\ \bibnamefont
  {Liu}}\ and\ \bibinfo {author} {\bibfnamefont {M.~D.}\ \bibnamefont
  {Newton}},\ }\bibfield  {title} {\bibinfo {title} {Reorganization energy for
  electron transfer at film-modified electrode surfaces: a dielectric continuum
  model},\ }\href@noop {} {\bibfield  {journal} {\bibinfo  {journal} {The
  Journal of Physical Chemistry}\ }\textbf {\bibinfo {volume} {98}},\ \bibinfo
  {pages} {7162} (\bibinfo {year} {1994})}\BibitemShut {NoStop}%
\bibitem [{\citenamefont {Medvedev}(2002)}]{medvedev_nonlocal_2002}%
  \BibitemOpen
  \bibfield  {author} {\bibinfo {author} {\bibfnamefont {I.~G.}\ \bibnamefont
  {Medvedev}},\ }\bibfield  {title} {\bibinfo {title} {Nonlocal effects and the
  overscreening effect in the kinetics of heterogeneous charge transfer
  reactions},\ }\href {https://doi.org/10.1023/A:1016864114447/METRICS}
  {\bibfield  {journal} {\bibinfo  {journal} {Russian Journal of
  Electrochemistry}\ }\textbf {\bibinfo {volume} {38}},\ \bibinfo {pages} {141}
  (\bibinfo {year} {2002})}\BibitemShut {NoStop}%
\bibitem [{\citenamefont {Dzhavakhidze}\ \emph {et~al.}(1987)\citenamefont
  {Dzhavakhidze}, \citenamefont {Kornyshev},\ and\ \citenamefont
  {Krishtalik}}]{dzhavakhidze_activation_1987}%
  \BibitemOpen
  \bibfield  {author} {\bibinfo {author} {\bibfnamefont {P.}~\bibnamefont
  {Dzhavakhidze}}, \bibinfo {author} {\bibfnamefont {A.}~\bibnamefont
  {Kornyshev}},\ and\ \bibinfo {author} {\bibfnamefont {L.}~\bibnamefont
  {Krishtalik}},\ }\bibfield  {title} {\bibinfo {title} {Activation energy of
  electrode reactions: the non-local effects},\ }\href@noop {} {\bibfield
  {journal} {\bibinfo  {journal} {Journal of electroanalytical chemistry and
  interfacial electrochemistry}\ }\textbf {\bibinfo {volume} {228}},\ \bibinfo
  {pages} {329} (\bibinfo {year} {1987})}\BibitemShut {NoStop}%
\bibitem [{\citenamefont {Medvedev}(2000)}]{medvedev_non-local_2000}%
  \BibitemOpen
  \bibfield  {author} {\bibinfo {author} {\bibfnamefont {I.~G.}\ \bibnamefont
  {Medvedev}},\ }\href {www.elsevier.nl/locate/jelechem} {\emph {\bibinfo
  {title} {Non-local effects in the kinetics of heterogeneous charge transfer
  reactions}}},\ \bibinfo {type} {Tech. Rep.}\ (\bibinfo {year} {2000})\
  \bibinfo {note} {volume: 481}\BibitemShut {NoStop}%
\bibitem [{\citenamefont {Kornyshev}\ \emph {et~al.}(1977)\citenamefont
  {Kornyshev}, \citenamefont {Rubinshtein},\ and\ \citenamefont
  {Vorotyntsev}}]{a_kornyshev__1977}%
  \BibitemOpen
  \bibfield  {author} {\bibinfo {author} {\bibfnamefont {A.}~\bibnamefont
  {Kornyshev}}, \bibinfo {author} {\bibfnamefont {A.}~\bibnamefont
  {Rubinshtein}},\ and\ \bibinfo {author} {\bibfnamefont {M.}~\bibnamefont
  {Vorotyntsev}},\ }\bibfield  {title} {\bibinfo {title} {Image potential near
  a dielectric--plasma-like medium interface},\ }\href@noop {} {\bibfield
  {journal} {\bibinfo  {journal} {physica status solidi (b)}\ }\textbf
  {\bibinfo {volume} {84}},\ \bibinfo {pages} {125} (\bibinfo {year}
  {1977})}\BibitemShut {NoStop}%
\bibitem [{\citenamefont {Fedorov}\ and\ \citenamefont
  {Kornyshev}(2014)}]{fedorov_ionic_2014}%
  \BibitemOpen
  \bibfield  {author} {\bibinfo {author} {\bibfnamefont {M.~V.}\ \bibnamefont
  {Fedorov}}\ and\ \bibinfo {author} {\bibfnamefont {A.~A.}\ \bibnamefont
  {Kornyshev}},\ }\bibfield  {title} {\bibinfo {title} {Ionic {Liquids} at
  {Electrified} {Interfaces}},\ }\href {https://doi.org/10.1021/cr400374x}
  {\bibfield  {journal} {\bibinfo  {journal} {Chemical Reviews}\ }\textbf
  {\bibinfo {volume} {114}},\ \bibinfo {pages} {2978} (\bibinfo {year}
  {2014})}\BibitemShut {NoStop}%
\bibitem [{\citenamefont {Kaiser}\ \emph {et~al.}(2017)\citenamefont {Kaiser},
  \citenamefont {Comtet}, \citenamefont {Nigu{\`e}s}, \citenamefont {Siria},
  \citenamefont {Coasne},\ and\ \citenamefont
  {Bocquet}}]{kaiser_electrostatic_2017}%
  \BibitemOpen
  \bibfield  {author} {\bibinfo {author} {\bibfnamefont {V.}~\bibnamefont
  {Kaiser}}, \bibinfo {author} {\bibfnamefont {J.}~\bibnamefont {Comtet}},
  \bibinfo {author} {\bibfnamefont {A.}~\bibnamefont {Nigu{\`e}s}}, \bibinfo
  {author} {\bibfnamefont {A.}~\bibnamefont {Siria}}, \bibinfo {author}
  {\bibfnamefont {B.}~\bibnamefont {Coasne}},\ and\ \bibinfo {author}
  {\bibfnamefont {L.}~\bibnamefont {Bocquet}},\ }\bibfield  {title} {\bibinfo
  {title} {Electrostatic interactions between ions near thomas--fermi
  substrates and the surface energy of ionic crystals at imperfect metals},\
  }\href@noop {} {\bibfield  {journal} {\bibinfo  {journal} {Faraday
  discussions}\ }\textbf {\bibinfo {volume} {199}},\ \bibinfo {pages} {129}
  (\bibinfo {year} {2017})}\BibitemShut {NoStop}%
\bibitem [{\citenamefont {Limaye}\ \emph {et~al.}(2020)\citenamefont {Limaye},
  \citenamefont {Ding},\ and\ \citenamefont
  {Willard}}]{limaye2020understanding}%
  \BibitemOpen
  \bibfield  {author} {\bibinfo {author} {\bibfnamefont {A.~M.}\ \bibnamefont
  {Limaye}}, \bibinfo {author} {\bibfnamefont {W.}~\bibnamefont {Ding}},\ and\
  \bibinfo {author} {\bibfnamefont {A.~P.}\ \bibnamefont {Willard}},\
  }\bibfield  {title} {\bibinfo {title} {Understanding attenuated solvent
  reorganization energies near electrode interfaces},\ }\href@noop {}
  {\bibfield  {journal} {\bibinfo  {journal} {The Journal of Chemical Physics}\
  }\textbf {\bibinfo {volume} {152}},\ \bibinfo {pages} {114706} (\bibinfo
  {year} {2020})}\BibitemShut {NoStop}%
\bibitem [{\citenamefont {Escalante}\ and\ \citenamefont
  {Limmer}()}]{SuppData}%
  \BibitemOpen
  \bibfield  {author} {\bibinfo {author} {\bibfnamefont {L.~C.}\ \bibnamefont
  {Escalante}}\ and\ \bibinfo {author} {\bibfnamefont {D.~T.}\ \bibnamefont
  {Limmer}},\ }\href {https://github.com/leo-coe/TBG_Electrochem} {\bibinfo
  {title} {Supplementary data,
  \url{https://github.com/leo-coe/TBG_Electrochem}}}\BibitemShut {NoStop}%
\end{thebibliography}%


\begin{thebibliography}{20}%
\makeatletter
\providecommand \@ifxundefined [1]{%
 \@ifx{#1\undefined}
}%
\providecommand \@ifnum [1]{%
 \ifnum #1\expandafter \@firstoftwo
 \else \expandafter \@secondoftwo
 \fi
}%
\providecommand \@ifx [1]{%
 \ifx #1\expandafter \@firstoftwo
 \else \expandafter \@secondoftwo
 \fi
}%
\providecommand \natexlab [1]{#1}%
\providecommand \enquote  [1]{``#1''}%
\providecommand \bibnamefont  [1]{#1}%
\providecommand \bibfnamefont [1]{#1}%
\providecommand \citenamefont [1]{#1}%
\providecommand \href@noop [0]{\@secondoftwo}%
\providecommand \href [0]{\begingroup \@sanitize@url \@href}%
\providecommand \@href[1]{\@@startlink{#1}\@@href}%
\providecommand \@@href[1]{\endgroup#1\@@endlink}%
\providecommand \@sanitize@url [0]{\catcode `\\12\catcode `\$12\catcode
  `\&12\catcode `\#12\catcode `\^12\catcode `\_12\catcode `\%12\relax}%
\providecommand \@@startlink[1]{}%
\providecommand \@@endlink[0]{}%
\providecommand \url  [0]{\begingroup\@sanitize@url \@url }%
\providecommand \@url [1]{\endgroup\@href {#1}{\urlprefix }}%
\providecommand \urlprefix  [0]{URL }%
\providecommand \Eprint [0]{\href }%
\providecommand \doibase [0]{https://doi.org/}%
\providecommand \selectlanguage [0]{\@gobble}%
\providecommand \bibinfo  [0]{\@secondoftwo}%
\providecommand \bibfield  [0]{\@secondoftwo}%
\providecommand \translation [1]{[#1]}%
\providecommand \BibitemOpen [0]{}%
\providecommand \bibitemStop [0]{}%
\providecommand \bibitemNoStop [0]{.\EOS\space}%
\providecommand \EOS [0]{\spacefactor3000\relax}%
\providecommand \BibitemShut  [1]{\csname bibitem#1\endcsname}%
\let\auto@bib@innerbib\@empty
\bibitem [{\citenamefont {Ahrens-Iwers}\ \emph {et~al.}(2022)\citenamefont
  {Ahrens-Iwers}, \citenamefont {Janssen}, \citenamefont {Tee},\ and\
  \citenamefont {Meißner}}]{ahrens-iwers_electrode_2022}%
  \BibitemOpen
  \bibfield  {author} {\bibinfo {author} {\bibfnamefont {L.~J.}\ \bibnamefont
  {Ahrens-Iwers}}, \bibinfo {author} {\bibfnamefont {M.}~\bibnamefont
  {Janssen}}, \bibinfo {author} {\bibfnamefont {S.~R.}\ \bibnamefont {Tee}},\
  and\ \bibinfo {author} {\bibfnamefont {R.~H.}\ \bibnamefont {Meißner}},\
  }\bibfield  {title} {\bibinfo {title} {{ELECTRODE}: {An} electrochemistry
  package for atomistic simulations},\ }\href
  {https://doi.org/10.1063/5.0099239/2841985} {\bibfield  {journal} {\bibinfo
  {journal} {Journal of Chemical Physics}\ }\textbf {\bibinfo {volume} {157}},\
  \bibinfo {pages} {84801} (\bibinfo {year} {2022})}\BibitemShut {NoStop}%
\bibitem [{\citenamefont {Thompson}\ \emph {et~al.}(2022)\citenamefont
  {Thompson}, \citenamefont {Aktulga}, \citenamefont {Berger}, \citenamefont
  {Bolintineanu}, \citenamefont {Brown}, \citenamefont {Crozier}, \citenamefont
  {in~'t Veld}, \citenamefont {Kohlmeyer}, \citenamefont {Moore}, \citenamefont
  {Nguyen}, \citenamefont {Shan}, \citenamefont {Stevens}, \citenamefont
  {Tranchida}, \citenamefont {Trott},\ and\ \citenamefont
  {Plimpton}}]{thompson_lammps_2022}%
  \BibitemOpen
  \bibfield  {author} {\bibinfo {author} {\bibfnamefont {A.~P.}\ \bibnamefont
  {Thompson}}, \bibinfo {author} {\bibfnamefont {H.~M.}\ \bibnamefont
  {Aktulga}}, \bibinfo {author} {\bibfnamefont {R.}~\bibnamefont {Berger}},
  \bibinfo {author} {\bibfnamefont {D.~S.}\ \bibnamefont {Bolintineanu}},
  \bibinfo {author} {\bibfnamefont {W.~M.}\ \bibnamefont {Brown}}, \bibinfo
  {author} {\bibfnamefont {P.~S.}\ \bibnamefont {Crozier}}, \bibinfo {author}
  {\bibfnamefont {P.~J.}\ \bibnamefont {in~'t Veld}}, \bibinfo {author}
  {\bibfnamefont {A.}~\bibnamefont {Kohlmeyer}}, \bibinfo {author}
  {\bibfnamefont {S.~G.}\ \bibnamefont {Moore}}, \bibinfo {author}
  {\bibfnamefont {T.~D.}\ \bibnamefont {Nguyen}}, \bibinfo {author}
  {\bibfnamefont {R.}~\bibnamefont {Shan}}, \bibinfo {author} {\bibfnamefont
  {M.~J.}\ \bibnamefont {Stevens}}, \bibinfo {author} {\bibfnamefont
  {J.}~\bibnamefont {Tranchida}}, \bibinfo {author} {\bibfnamefont
  {C.}~\bibnamefont {Trott}},\ and\ \bibinfo {author} {\bibfnamefont {S.~J.}\
  \bibnamefont {Plimpton}},\ }\bibfield  {title} {\bibinfo {title} {{LAMMPS} -
  a flexible simulation tool for particle-based materials modeling at the
  atomic, meso, and continuum scales},\ }\href
  {https://doi.org/10.1016/j.cpc.2021.108171} {\bibfield  {journal} {\bibinfo
  {journal} {Computer Physics Communications}\ }\textbf {\bibinfo {volume}
  {271}},\ \bibinfo {pages} {108171} (\bibinfo {year} {2022})}\BibitemShut
  {NoStop}%
\bibitem [{\citenamefont {Ferrario}\ \emph {et~al.}(2006)\citenamefont
  {Ferrario}, \citenamefont {Ciccotti},\ and\ \citenamefont
  {Binder}}]{ferrario_computer_2006}%
  \BibitemOpen
  \bibinfo {editor} {\bibfnamefont {M.}~\bibnamefont {Ferrario}}, \bibinfo
  {editor} {\bibfnamefont {G.}~\bibnamefont {Ciccotti}},\ and\ \bibinfo
  {editor} {\bibfnamefont {K.}~\bibnamefont {Binder}},\ eds.,\ \href
  {https://doi.org/10.1007/3-540-35284-8} {\emph {\bibinfo {title} {Computer
  {Simulations} in {Condensed} {Matter} {Systems}: {From} {Materials} to
  {Chemical} {Biology} {Volume} 2}}},\ Vol.\ \bibinfo {volume} {704}\ (\bibinfo
   {publisher} {Springer Berlin Heidelberg},\ \bibinfo {address} {Berlin,
  Heidelberg},\ \bibinfo {year} {2006})\BibitemShut {NoStop}%
\bibitem [{\citenamefont {Medvedev}(2002)}]{medvedev_nonlocal_2002}%
  \BibitemOpen
  \bibfield  {author} {\bibinfo {author} {\bibfnamefont {I.~G.}\ \bibnamefont
  {Medvedev}},\ }\bibfield  {title} {\bibinfo {title} {Nonlocal effects and the
  overscreening effect in the kinetics of heterogeneous charge transfer
  reactions},\ }\href {https://doi.org/10.1023/A:1016864114447/METRICS}
  {\bibfield  {journal} {\bibinfo  {journal} {Russian Journal of
  Electrochemistry}\ }\textbf {\bibinfo {volume} {38}},\ \bibinfo {pages} {141}
  (\bibinfo {year} {2002})}\BibitemShut {NoStop}%
\bibitem [{\citenamefont {Dzhavakhidze}\ \emph {et~al.}(1987)\citenamefont
  {Dzhavakhidze}, \citenamefont {Kornyshev},\ and\ \citenamefont
  {Krishtalik}}]{dzhavakhidze_activation_1987}%
  \BibitemOpen
  \bibfield  {author} {\bibinfo {author} {\bibfnamefont {P.}~\bibnamefont
  {Dzhavakhidze}}, \bibinfo {author} {\bibfnamefont {A.}~\bibnamefont
  {Kornyshev}},\ and\ \bibinfo {author} {\bibfnamefont {L.}~\bibnamefont
  {Krishtalik}},\ }\bibfield  {title} {\bibinfo {title} {Activation energy of
  electrode reactions: the non-local effects},\ }\href@noop {} {\bibfield
  {journal} {\bibinfo  {journal} {Journal of electroanalytical chemistry and
  interfacial electrochemistry}\ }\textbf {\bibinfo {volume} {228}},\ \bibinfo
  {pages} {329} (\bibinfo {year} {1987})}\BibitemShut {NoStop}%
\bibitem [{\citenamefont {Medvedev}(2000)}]{medvedev_non-local_2000-1}%
  \BibitemOpen
  \bibfield  {author} {\bibinfo {author} {\bibfnamefont {I.}~\bibnamefont
  {Medvedev}},\ }\bibfield  {title} {\bibinfo {title} {Non-local effects in the
  kinetics of heterogeneous charge transfer reactions},\ }\href@noop {}
  {\bibfield  {journal} {\bibinfo  {journal} {Journal of Electroanalytical
  Chemistry}\ }\textbf {\bibinfo {volume} {481}},\ \bibinfo {pages} {215}
  (\bibinfo {year} {2000})}\BibitemShut {NoStop}%
\bibitem [{\citenamefont {Kornyshev}\ \emph {et~al.}(1977)\citenamefont
  {Kornyshev}, \citenamefont {Rubinshtein},\ and\ \citenamefont
  {Vorotyntsev}}]{a_kornyshev__1977}%
  \BibitemOpen
  \bibfield  {author} {\bibinfo {author} {\bibfnamefont {A.}~\bibnamefont
  {Kornyshev}}, \bibinfo {author} {\bibfnamefont {A.}~\bibnamefont
  {Rubinshtein}},\ and\ \bibinfo {author} {\bibfnamefont {M.}~\bibnamefont
  {Vorotyntsev}},\ }\bibfield  {title} {\bibinfo {title} {Image potential near
  a dielectric--plasma-like medium interface},\ }\href@noop {} {\bibfield
  {journal} {\bibinfo  {journal} {physica status solidi (b)}\ }\textbf
  {\bibinfo {volume} {84}},\ \bibinfo {pages} {125} (\bibinfo {year}
  {1977})}\BibitemShut {NoStop}%
\bibitem [{\citenamefont {Fedorov}\ and\ \citenamefont
  {Kornyshev}(2014)}]{fedorov_ionic_2014}%
  \BibitemOpen
  \bibfield  {author} {\bibinfo {author} {\bibfnamefont {M.~V.}\ \bibnamefont
  {Fedorov}}\ and\ \bibinfo {author} {\bibfnamefont {A.~A.}\ \bibnamefont
  {Kornyshev}},\ }\bibfield  {title} {\bibinfo {title} {Ionic {Liquids} at
  {Electrified} {Interfaces}},\ }\href {https://doi.org/10.1021/cr400374x}
  {\bibfield  {journal} {\bibinfo  {journal} {Chemical Reviews}\ }\textbf
  {\bibinfo {volume} {114}},\ \bibinfo {pages} {2978} (\bibinfo {year}
  {2014})}\BibitemShut {NoStop}%
\bibitem [{\citenamefont {Kaiser}\ \emph {et~al.}(2017)\citenamefont {Kaiser},
  \citenamefont {Comtet}, \citenamefont {Nigu{\`e}s}, \citenamefont {Siria},
  \citenamefont {Coasne},\ and\ \citenamefont
  {Bocquet}}]{kaiser_electrostatic_2017}%
  \BibitemOpen
  \bibfield  {author} {\bibinfo {author} {\bibfnamefont {V.}~\bibnamefont
  {Kaiser}}, \bibinfo {author} {\bibfnamefont {J.}~\bibnamefont {Comtet}},
  \bibinfo {author} {\bibfnamefont {A.}~\bibnamefont {Nigu{\`e}s}}, \bibinfo
  {author} {\bibfnamefont {A.}~\bibnamefont {Siria}}, \bibinfo {author}
  {\bibfnamefont {B.}~\bibnamefont {Coasne}},\ and\ \bibinfo {author}
  {\bibfnamefont {L.}~\bibnamefont {Bocquet}},\ }\bibfield  {title} {\bibinfo
  {title} {Electrostatic interactions between ions near thomas--fermi
  substrates and the surface energy of ionic crystals at imperfect metals},\
  }\href@noop {} {\bibfield  {journal} {\bibinfo  {journal} {Faraday
  discussions}\ }\textbf {\bibinfo {volume} {199}},\ \bibinfo {pages} {129}
  (\bibinfo {year} {2017})}\BibitemShut {NoStop}%
\bibitem [{\citenamefont {Noori}\ \emph {et~al.}(2019)\citenamefont {Noori},
  \citenamefont {Cheng}, \citenamefont {Xuan},\ and\ \citenamefont
  {Quek}}]{noori_dielectric_2019}%
  \BibitemOpen
  \bibfield  {author} {\bibinfo {author} {\bibfnamefont {K.}~\bibnamefont
  {Noori}}, \bibinfo {author} {\bibfnamefont {N.~L.~Q.}\ \bibnamefont {Cheng}},
  \bibinfo {author} {\bibfnamefont {F.}~\bibnamefont {Xuan}},\ and\ \bibinfo
  {author} {\bibfnamefont {S.~Y.}\ \bibnamefont {Quek}},\ }\bibfield  {title}
  {\bibinfo {title} {Dielectric screening by {2D} substrates},\ }\href
  {https://doi.org/10.1088/2053-1583/ab1e06} {\bibfield  {journal} {\bibinfo
  {journal} {2D Materials}\ }\textbf {\bibinfo {volume} {6}},\ \bibinfo {pages}
  {035036} (\bibinfo {year} {2019})}\BibitemShut {NoStop}%
\bibitem [{\citenamefont {Ando}\ \emph {et~al.}(1982)\citenamefont {Ando},
  \citenamefont {Fowler},\ and\ \citenamefont {Stern}}]{ando_electronic_1982}%
  \BibitemOpen
  \bibfield  {author} {\bibinfo {author} {\bibfnamefont {T.}~\bibnamefont
  {Ando}}, \bibinfo {author} {\bibfnamefont {A.~B.}\ \bibnamefont {Fowler}},\
  and\ \bibinfo {author} {\bibfnamefont {F.}~\bibnamefont {Stern}},\ }\bibfield
   {title} {\bibinfo {title} {Electronic properties of two-dimensional
  systems},\ }\href {https://doi.org/10.1103/RevModPhys.54.437} {\bibfield
  {journal} {\bibinfo  {journal} {Reviews of Modern Physics}\ }\textbf
  {\bibinfo {volume} {54}},\ \bibinfo {pages} {437} (\bibinfo {year}
  {1982})}\BibitemShut {NoStop}%
\bibitem [{\citenamefont {Parhizgar}\ \emph {et~al.}(2017)\citenamefont
  {Parhizgar}, \citenamefont {Qaiumzadeh},\ and\ \citenamefont
  {Asgari}}]{parhizgar2017quantum}%
  \BibitemOpen
  \bibfield  {author} {\bibinfo {author} {\bibfnamefont {F.}~\bibnamefont
  {Parhizgar}}, \bibinfo {author} {\bibfnamefont {A.}~\bibnamefont
  {Qaiumzadeh}},\ and\ \bibinfo {author} {\bibfnamefont {R.}~\bibnamefont
  {Asgari}},\ }\bibfield  {title} {\bibinfo {title} {Quantum capacitance of
  double-layer graphene},\ }\href@noop {} {\bibfield  {journal} {\bibinfo
  {journal} {Physical Review B}\ }\textbf {\bibinfo {volume} {96}},\ \bibinfo
  {pages} {075447} (\bibinfo {year} {2017})}\BibitemShut {NoStop}%
\bibitem [{\citenamefont {Yu}\ \emph {et~al.}(2022)\citenamefont {Yu},
  \citenamefont {Zhang}, \citenamefont {Parks}, \citenamefont {Babar},
  \citenamefont {Carr}, \citenamefont {Craig}, \citenamefont {Van~Winkle},
  \citenamefont {Lyssenko}, \citenamefont {Taniguchi}, \citenamefont
  {Watanabe}, \citenamefont {Viswanathan},\ and\ \citenamefont
  {Bediako}}]{yu_tunable_2022}%
  \BibitemOpen
  \bibfield  {author} {\bibinfo {author} {\bibfnamefont {Y.}~\bibnamefont
  {Yu}}, \bibinfo {author} {\bibfnamefont {K.}~\bibnamefont {Zhang}}, \bibinfo
  {author} {\bibfnamefont {H.}~\bibnamefont {Parks}}, \bibinfo {author}
  {\bibfnamefont {M.}~\bibnamefont {Babar}}, \bibinfo {author} {\bibfnamefont
  {S.}~\bibnamefont {Carr}}, \bibinfo {author} {\bibfnamefont {I.~M.}\
  \bibnamefont {Craig}}, \bibinfo {author} {\bibfnamefont {M.}~\bibnamefont
  {Van~Winkle}}, \bibinfo {author} {\bibfnamefont {A.}~\bibnamefont
  {Lyssenko}}, \bibinfo {author} {\bibfnamefont {T.}~\bibnamefont {Taniguchi}},
  \bibinfo {author} {\bibfnamefont {K.}~\bibnamefont {Watanabe}}, \bibinfo
  {author} {\bibfnamefont {V.}~\bibnamefont {Viswanathan}},\ and\ \bibinfo
  {author} {\bibfnamefont {D.~K.}\ \bibnamefont {Bediako}},\ }\bibfield
  {title} {\bibinfo {title} {Tunable angle-dependent electrochemistry at
  twisted bilayer graphene with moiré flat bands},\ }\href
  {https://doi.org/10.1038/s41557-021-00865-1} {\bibfield  {journal} {\bibinfo
  {journal} {Nature Chemistry}\ }\textbf {\bibinfo {volume} {14}},\ \bibinfo
  {pages} {267} (\bibinfo {year} {2022})}\BibitemShut {NoStop}%
\bibitem [{\citenamefont {Wong}\ \emph {et~al.}(2020)\citenamefont {Wong},
  \citenamefont {Nuckolls}, \citenamefont {Oh}, \citenamefont {Lian},
  \citenamefont {Xie}, \citenamefont {Jeon}, \citenamefont {Watanabe},
  \citenamefont {Taniguchi}, \citenamefont {Bernevig},\ and\ \citenamefont
  {Yazdani}}]{wong_cascade_2020}%
  \BibitemOpen
  \bibfield  {author} {\bibinfo {author} {\bibfnamefont {D.}~\bibnamefont
  {Wong}}, \bibinfo {author} {\bibfnamefont {K.~P.}\ \bibnamefont {Nuckolls}},
  \bibinfo {author} {\bibfnamefont {M.}~\bibnamefont {Oh}}, \bibinfo {author}
  {\bibfnamefont {B.}~\bibnamefont {Lian}}, \bibinfo {author} {\bibfnamefont
  {Y.}~\bibnamefont {Xie}}, \bibinfo {author} {\bibfnamefont {S.}~\bibnamefont
  {Jeon}}, \bibinfo {author} {\bibfnamefont {K.}~\bibnamefont {Watanabe}},
  \bibinfo {author} {\bibfnamefont {T.}~\bibnamefont {Taniguchi}}, \bibinfo
  {author} {\bibfnamefont {B.~A.}\ \bibnamefont {Bernevig}},\ and\ \bibinfo
  {author} {\bibfnamefont {A.}~\bibnamefont {Yazdani}},\ }\bibfield  {title}
  {\bibinfo {title} {Cascade of electronic transitions in magic-angle twisted
  bilayer graphene},\ }\href {https://doi.org/10.1038/s41586-020-2339-0}
  {\bibfield  {journal} {\bibinfo  {journal} {Nature}\ }\textbf {\bibinfo
  {volume} {582}},\ \bibinfo {pages} {198} (\bibinfo {year}
  {2020})}\BibitemShut {NoStop}%
\bibitem [{\citenamefont {Rozhkov}\ \emph {et~al.}(2016)\citenamefont
  {Rozhkov}, \citenamefont {Sboychakov}, \citenamefont {Rakhmanov},\ and\
  \citenamefont {Nori}}]{rozhkov_electronic_nodate}%
  \BibitemOpen
  \bibfield  {author} {\bibinfo {author} {\bibfnamefont {A.~V.}\ \bibnamefont
  {Rozhkov}}, \bibinfo {author} {\bibfnamefont {A.}~\bibnamefont {Sboychakov}},
  \bibinfo {author} {\bibfnamefont {A.}~\bibnamefont {Rakhmanov}},\ and\
  \bibinfo {author} {\bibfnamefont {F.}~\bibnamefont {Nori}},\ }\bibfield
  {title} {\bibinfo {title} {Electronic properties of graphene-based bilayer
  systems},\ }\href@noop {} {\bibfield  {journal} {\bibinfo  {journal} {Physics
  Reports}\ }\textbf {\bibinfo {volume} {648}},\ \bibinfo {pages} {1} (\bibinfo
  {year} {2016})}\BibitemShut {NoStop}%
\bibitem [{\citenamefont {Dos~Santos}\ \emph {et~al.}(2007)\citenamefont
  {Dos~Santos}, \citenamefont {Peres},\ and\ \citenamefont
  {Neto}}]{lopes_graphene_2007}%
  \BibitemOpen
  \bibfield  {author} {\bibinfo {author} {\bibfnamefont {J.~L.}\ \bibnamefont
  {Dos~Santos}}, \bibinfo {author} {\bibfnamefont {N.}~\bibnamefont {Peres}},\
  and\ \bibinfo {author} {\bibfnamefont {A.~C.}\ \bibnamefont {Neto}},\
  }\bibfield  {title} {\bibinfo {title} {Graphene bilayer with a twist:
  Electronic structure},\ }\href@noop {} {\bibfield  {journal} {\bibinfo
  {journal} {Physical review letters}\ }\textbf {\bibinfo {volume} {99}},\
  \bibinfo {pages} {256802} (\bibinfo {year} {2007})}\BibitemShut {NoStop}%
\bibitem [{\citenamefont {Bistritzer}\ and\ \citenamefont
  {MacDonald}(2011)}]{bistritzer_moire_2011}%
  \BibitemOpen
  \bibfield  {author} {\bibinfo {author} {\bibfnamefont {R.}~\bibnamefont
  {Bistritzer}}\ and\ \bibinfo {author} {\bibfnamefont {A.~H.}\ \bibnamefont
  {MacDonald}},\ }\bibfield  {title} {\bibinfo {title} {Moiré bands in twisted
  double-layer graphene},\ }\href {https://doi.org/10.1073/PNAS.1108174108}
  {\bibfield  {journal} {\bibinfo  {journal} {Proceedings of the National
  Academy of Sciences of the United States of America}\ }\textbf {\bibinfo
  {volume} {108}},\ \bibinfo {pages} {12233} (\bibinfo {year}
  {2011})}\BibitemShut {NoStop}%
\bibitem [{\citenamefont {Carr}\ \emph {et~al.}(2019)\citenamefont {Carr},
  \citenamefont {Fang}, \citenamefont {Zhu},\ and\ \citenamefont
  {Kaxiras}}]{carr2019exact}%
  \BibitemOpen
  \bibfield  {author} {\bibinfo {author} {\bibfnamefont {S.}~\bibnamefont
  {Carr}}, \bibinfo {author} {\bibfnamefont {S.}~\bibnamefont {Fang}}, \bibinfo
  {author} {\bibfnamefont {Z.}~\bibnamefont {Zhu}},\ and\ \bibinfo {author}
  {\bibfnamefont {E.}~\bibnamefont {Kaxiras}},\ }\bibfield  {title} {\bibinfo
  {title} {Exact continuum model for low-energy electronic states of twisted
  bilayer graphene},\ }\href@noop {} {\bibfield  {journal} {\bibinfo  {journal}
  {Physical Review Research}\ }\textbf {\bibinfo {volume} {1}},\ \bibinfo
  {pages} {013001} (\bibinfo {year} {2019})}\BibitemShut {NoStop}%
\bibitem [{\citenamefont {Carr}\ \emph {et~al.}(2020)\citenamefont {Carr},
  \citenamefont {Fang},\ and\ \citenamefont {Kaxiras}}]{carr2020electronic}%
  \BibitemOpen
  \bibfield  {author} {\bibinfo {author} {\bibfnamefont {S.}~\bibnamefont
  {Carr}}, \bibinfo {author} {\bibfnamefont {S.}~\bibnamefont {Fang}},\ and\
  \bibinfo {author} {\bibfnamefont {E.}~\bibnamefont {Kaxiras}},\ }\bibfield
  {title} {\bibinfo {title} {Electronic-structure methods for twisted moir{\'e}
  layers},\ }\href@noop {} {\bibfield  {journal} {\bibinfo  {journal} {Nature
  Reviews Materials}\ }\textbf {\bibinfo {volume} {5}},\ \bibinfo {pages} {748}
  (\bibinfo {year} {2020})}\BibitemShut {NoStop}%
\bibitem [{\citenamefont {Fang}\ and\ \citenamefont
  {Kaxiras}(2016)}]{fang2016electronic}%
  \BibitemOpen
  \bibfield  {author} {\bibinfo {author} {\bibfnamefont {S.}~\bibnamefont
  {Fang}}\ and\ \bibinfo {author} {\bibfnamefont {E.}~\bibnamefont {Kaxiras}},\
  }\bibfield  {title} {\bibinfo {title} {Electronic structure theory of weakly
  interacting bilayers},\ }\href@noop {} {\bibfield  {journal} {\bibinfo
  {journal} {Physical Review B}\ }\textbf {\bibinfo {volume} {93}},\ \bibinfo
  {pages} {235153} (\bibinfo {year} {2016})}\BibitemShut {NoStop}%
\end{thebibliography}%

\end{document}


\preprint{1}

\title{Supplemental Material for ``Microscopic origin of twist-dependent electron transfer rate in bilayer graphene"}

\author{Leonardo Coello Escalante}
\affiliation{%
 Department of Chemistry, University of California, Berkeley, CA 94720, USA
}%

\author{David T. Limmer}
 \email{dlimmer@berkeley.edu}
\affiliation{%
 Department of Chemistry, University of California, Berkeley, CA 94720, USA
}%
\affiliation{%
 Kavli Energy NanoScience Institute, Berkeley, CA 94720, USA
}%
\affiliation{%
 MSD, Lawrence Berkeley National Laboratory, Berkeley, CA 94720, USA
}%
\affiliation{%
 CSD, Lawrence Berkeley, National Laboratory, Berkeley, CA 94720, USA
}%

\date{\today}


\maketitle

\section{Simulation Details}

Constant potential molecular dynamics simulations were run, and energy gap statistics were sampled via thermodynamic integration for three model outer sphere redox couples ($\ch{Fe^3+}/\ch{Fe^2+}$, $\ch{K^+}/\ch{K^0}$ and $\ch{Cl^0}/\ch{Cl^-}$), and four different values of screening length ($\ltf = 0 \ \mathrm{\AA}, 2 \ \mathrm{\AA}, 5 \ \mathrm{\AA}$ and $\ltf \to \infty$), at zero applied potential. All simulations were carried out using the ELECTRODE package \cite{ahrens-iwers_electrode_2022} implemented in LAMMPS \cite{thompson_lammps_2022}. Simulations were done using the finite field method, allowing Coulomb interactions to be computed using the regular particle-mesh approach to Ewald summation with full three dimensional periodicity. The tolerance in the Ewald sum was set to $1\times10^{-7} \ \mathrm{kcal/mol}$. Dynamics were evolved with a Langevin thermostat at 298 K with a damping parameter of $2 \mathrm{ps}$, using a timestep of $1 \mathrm{fs}$.
In constant potential simulations, the fluctuating charge density in the electrode is modeled with atom-centered Gaussian charge distributions, with a parameter $\eta$ setting the extent of the local distribution
\begin{equation}
    \label{elec_charge_density}
    \rho_{\mathrm{elec}}(\textbf{r}) = \sum_{j=1}^M q_j\eta^3\pi^{-3/2}e^{-\eta^2|\textbf{r}-\textbf{r}_j|^2}
\end{equation}
and for all of our simulations, the parameter $\eta$ was set to $1.979 \ \mathrm{\AA} ^{-1}$

An alternative to the finite field method is to employ two distinct electrodes and use two-dimensional Ewald sums. To verify consistency in the calculation of electrostatic interactions, we have run shorter tests and verified that the Marcus free energy surfaces that result from using the finite field method are consistent with simulations using 2D Ewald.

We used the same force field for oxidized and reduced species, so that the energy gap only contains contributions from changes in electrostatic interactions. This means that the biased ensembles used for thermodynamic integration can be constructed simply by assigning a fractional charge value to the redox species, in-between the oxidized an reduced forms
\begin{equation}
    q_\eta = q_D + \eta\delta q
\end{equation}
where $\delta q = q_A - q_D$ that always equal to $\pm 1 e$ in the systems studied. We ran simulations for nine different biasing windows ($\eta \in \{-0.2,0.0,0.2,0.4,0.5,0.6,0.8,1.0,1.2\}$) in order to gain information about the extreme tails of the distribution. Each biasing window was first equilibrated for 1 ns, and then run for an additional 2 ns, recording the system's configuration every 0.5 ps and evaluating the energies in the oxidized and reduced states, with the difference being the energy gap $\Delta E$. For each state and each configuration, the charges on the electrode are relaxed subject to the constant potential constraint. 

\begin{figure*}[t]
            \includegraphics[width=17cm]{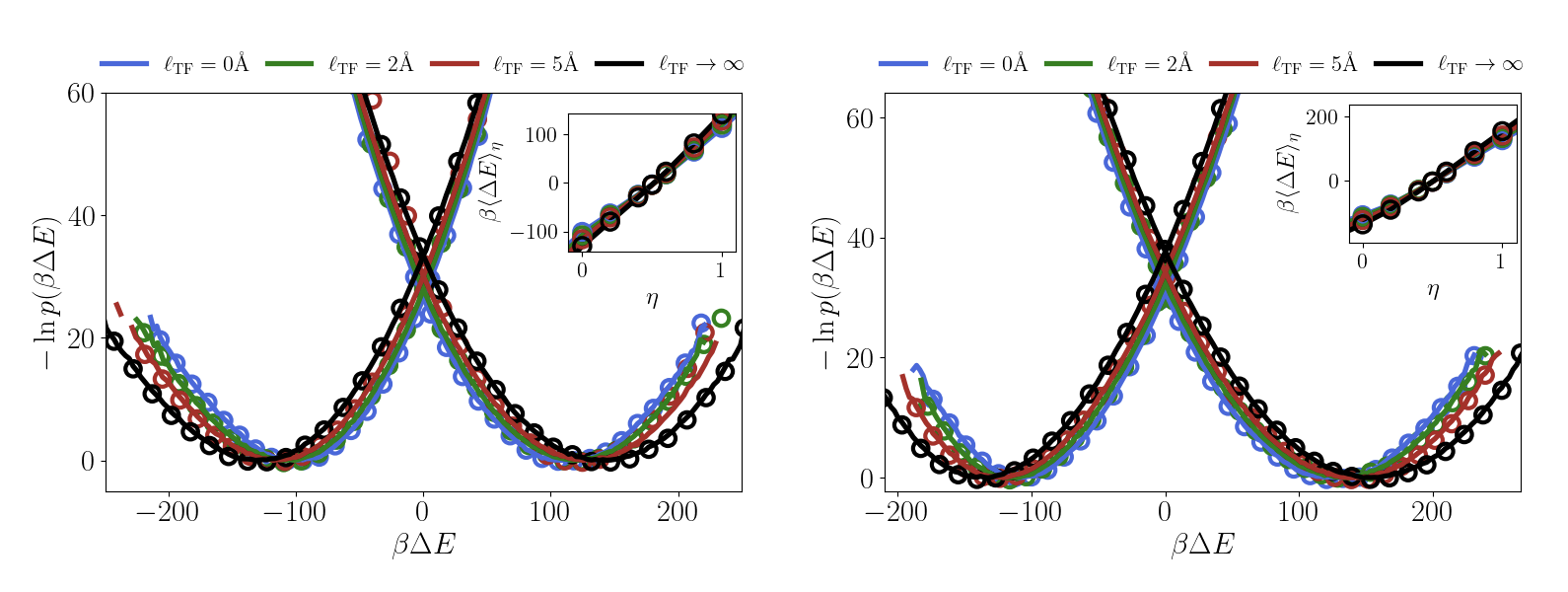}
    \caption{Free energy surfaces of electron transfer for the $\ch{K}^+/\ch{K}$ (left) and $\ch{Cl}/\ch{Cl}^-$ (right) redox couples. Solid lines correspond to the surfaces obtained from thermodynamic integration, open circles are the surfaces constructed from the linear dependence of the two curves (see Eq. \ref{lin_dep}). Inset shows the average value of the energy gap in the biased ensembles as a function of biasing parameter}
    \label{fig:Marcus_curves}
\end{figure*}

\section{Marcus Curves}
The electron transfer free energy surfaces were constructed from the biased trajectory data using MBAR. The change in free energy was computed both from MBAR, and through the following relation derived from thermodynamic integration
\begin{equation}
        \label{driving_thermoint}
	\Delta F = \int_0^1 \langle\Delta E\rangle_\eta d\eta
\end{equation}
and checked against the linear response relation
\begin{equation}
    \label{driving_linresp}
    \Delta F \approx \frac{1}{2}\left(\langle\Delta E\rangle_A + \langle\Delta E\rangle_D\right) \, .
\end{equation}
where the subscripts $A$ and $D$ refer to $\eta=0$ and $\eta=1$, respectively. The distribution of energy gaps for the $\ch{K}^+/\ch{K}$ and $\ch{Cl}/\ch{Cl}^-$ redox couples, not included in the main text, are shown in Fig. \ref{fig:Marcus_curves}. The free energy as a function of energy gap obeys the relationship
\begin{equation}
\label{lin_dep}
    F_D(\Delta E) = F_A(\Delta E) + \Delta E -\Delta F
\end{equation}
which we have verified by independently computing the free energy within the donor and acceptor states and relating them to each other through this relationship, also shown in Fig. \ref{fig:Marcus_curves}.

The values of free energy of activation, reorganization energy and driving force for all the systems are shown in Table \ref{tab:act_reorg}. In all cases, there is a difference in activation energy of roughly $7 \ \kB T$ between $\ltf = 0 \ \mathrm{\AA}$ and $\ltf \to \infty$, which translates to a difference in rate constants of up to 3 orders of magnitude. 
The Marcus curves shown in Fig. \ref{fig:Marcus_curves} were all displaced by a constant horizontal shift in order for their intersection to be located at $\Delta E = 0$. This is due to a well-known limitation of classical simulation studies of electrochemical reactions, related to the fact that the simulation does not have access to information about the true value of the work function of the electrode. 

Linear response is observed to be valid in all cases. This can be confirmed through several metrics, such as the agreement in the value of the several definitions of the reorganization energy \cite{ferrario_computer_2006}, and the almost exactly linear relationship of $\langle\Delta E\rangle_\eta$ as a function of $\eta$ shown in the insets of Fig. \ref{fig:Marcus_curves}. Our thermodynamic integration approach provides us with a way of computing the reorganization energy by computing the slope of the following linear relationship
\begin{equation}
    \label{reorg_thermoint}
    \langle\Delta E\rangle_\eta \approx \langle\Delta E\rangle_D -2\lambda\eta
\end{equation}
where $\lambda$ is the reorganization energy. Furthermore, the reorganization energy is defined in three additional ways that should all be equivalent within linear response,\begin{align}
    \label{reorg_stokes}
    \lambda &\approx F_{D}(\langle\Delta E\rangle_{A}) - F_{D}(\langle\Delta E\rangle_{D}) \\
    \label{reorg_linresp}
    &\approx \frac{1}{2}\left(\langle\Delta E\rangle_A - \langle\Delta E\rangle_D\right) \\
    \label{reorg_fluct}
    &\approx \frac{\beta}{2} \langle(\delta(\Delta E))^2\rangle_A \approx \frac{\beta}{2} \langle(\delta(\Delta E))^2\rangle_D 
\end{align}
Table \ref{tab:reorg_comp} compares the values obtained from all these definitions. We can see that in most instances there is excellent agreement between all definitions, confirming that the dielectric response of our systems is approximately linear.

\begin{table}[]
    \centering
    \begin{tabular}{cccccc}
        $\ltf / \mathrm{\AA}$ & $\beta F^{\ddagger}(\ltf)$ & $\beta \Delta F^{\ddagger}(\ltf)$ & $\beta \lambda(\ltf)$ & $\beta \Delta \lambda(\ltf)$ & $\beta \Delta \varepsilon$ \\
        \hline
        \hline
        $\ch{Fe^{3+}}/\ch{Fe^{2+}}$\\
        \hline
         0 & 21.2 & 0 & 86.2 & 0 & $\sim$ 0\\
         2 & 22.6 & 1.4 & 92.2 & 6.0 & $\sim$ 0\\
         5 & 24.7 & 3.5 & 99.9 & 13.7 & $\sim$ 0\\
         $\infty$ & 28.2 & 7.0 & 113.3 & 27.1 & 0.04\\
        \hline
        \hline
        $\ch{K^{+}}/\ch{K^{0}}$ \\
        \hline
         0 & 26.1 & 0 & 102.5 & 0 & -0.24\\
         2 & 27.2 &  1.1 & 108.5 & 6.0 & $\sim$ 0\\
         5 & 29.5 & 3.4 & 116.8 & 14.3 & 0.008\\
         $\infty$ & 32.9 & 6.8 & 130.9 & 28.4 & $\sim$ 0\\
         \hline
         \hline
        $\ch{Cl^{0}}/\ch{Cl^{-}}$ \\
        \hline
         0 & 30.8 & 0 & 115.6 & 0 & -0.02\\
         2 & 32.4 & 1.6 & 121.6 & 6.0 & $\sim$ 0\\
         5 & 34.2 & 3.4 & 129.7 & 14.1 & $\sim$ 0\\
         $\infty$ & 37.6 & 6.8 & 143.4 & 27.8 & $\sim$ 0\\
        \hline
        \hline
    \end{tabular}
    \caption{Free energy of activation, reorganization energy and thermodynamic driving force of the three redox systems studied. Also included is the difference in $F^{\ddagger}$ and $\lambda$ with respect to $\ltf = 0$}
    \label{tab:act_reorg}
\end{table}

\begin{table}[]
    \centering
    \begin{tabular}{cccccc}
        $\ltf / \mathrm{\AA}$ & $\beta \lambda_\mathrm{Stokes}$\footnote{Eq. \ref{reorg_stokes}} & $\beta \lambda_\mathrm{TE}$\footnote{Eq. \ref{reorg_thermoint}} & $\beta \lambda_\mathrm{LR}$ \footnote{Eq. \ref{reorg_linresp}} & $\beta \lambda_\mathrm{fluct}$ \footnote{Eq. \ref{reorg_fluct}(left surface/ right surface)} \\
        \hline
        \hline
        $\ch{Fe^{3+}}/\ch{Fe^{2+}}$\\
        \hline
         0 & 86.3 & 86.2 & 86.4 & 84.9/85.0 \\
         2 & 83.0 & 92.2 & 83.4 & 89.5/89.9 \\
         5 & 94.2 & 99.9 & 94.9 & 96.9/97.3 \\
         $\infty$ & 106.8 & 113.3 & 107.6 & 111.1/111.2 \\
        \hline
        \hline
        $\ch{K^{+}}/\ch{K^{0}}$ \\
        \hline
         0 & 108.0 & 102.5 & 108.2 & 112.6/111.5 \\
         2 & 120.9 & 108.5 & 121.2 & 117.8/117.4 \\
         5 & 130.9 & 116.8 & 124.9 & 125.5/125.0 \\
         $\infty$ & 137.7 & 130.9 & 138.0 & 138.9/138.4 \\
         \hline
         \hline
        $\ch{Cl^{0}}/\ch{Cl^{-}}$ \\
        \hline
         0 & 109.8 & 115.6 & 109.9 & 121.2/120.4 \\
         2 & 115.9 & 121.6 & 116.2 & 128.4/126.8 \\
         5 & 120.8 & 129.7 & 121.1 & 135.7/134.2 \\
         $\infty$ & 132.0 & 143.4 & 132.2 & 149.3/148.2 \\
        \hline
        \hline
    \end{tabular}
    \caption{Comparison of the various definitions of reorganization energy for all the redox couples and screening lengths studied}
    \label{tab:reorg_comp}
\end{table}


\section{Dielectric Continuum Theory Estimate of $\lambda(\ell_\mathrm{TF})$}

In this section we derive the dielectric continuum equations that describe the screening-dependent reorganization energy.  Our approach is an adaptation of a previous theory \cite{medvedev_nonlocal_2002, dzhavakhidze_activation_1987, medvedev_non-local_2000-1, a_kornyshev__1977, fedorov_ionic_2014, kaiser_electrostatic_2017}, though we provide a formulation that we believe to be amenable to intuitive interpretation, especially in connection to simulation studies of electron transfer. 

Consider a point charge $q$ in a medium with dielectric constant $\epsilon^{(\text{sol})}$ located at $\textbf{r}_0$, a distance $z_0>0$ away from a planar interface (located at $z=0$) with an electrode characterized by screening length $\ltf$ and a static dielectric constant $\epsilon^{(\text{el})}$. Poisson's equation at the boundary may be solved making use of Fourier-Bessel transforms. The resulting transform of the potential is expressible in cylindrical coordinates in the solution, $z>0$,\cite{a_kornyshev__1977}
\begin{align}
    \label{potential_sol}
        \phi_{\mathrm{sol}}&(\rho,\theta,z) = \frac{q}{\epsilon^{(\mathrm{sol})}} \Bigg[ \int_0^\infty J_0(k\rho) e^{-k|z-z_0|} dk \\
         &+ \int_0^\infty  J_0(k\rho) e^{-k(z+z_0)} \frac{\epsilon^{(\mathrm{sol})}k-\sqrt{k^2+\ltf^{-2}}}{\epsilon^{(\mathrm{sol})}k+\sqrt{k^2+\ltf^{-2}}} dk \Bigg] \notag
\end{align}
while at the electrode, $z=0$, 
\begin{align}
    \label{potential_elec}
        \phi_{\mathrm{el}}(\rho,\theta,z) = q \int_0^\infty& \Bigg( J_0(k\rho) e^{-kz_0}e^{-z_0\sqrt{k^2+\ltf^{-2}}} \\ 
        &\times \frac{2k}{\epsilon^{(\mathrm{sol})}k+\sqrt{k^2+\ltf^{-2}}} \Bigg) dk \notag
\end{align}
where $\rho = \sqrt{x^2 + y^2}$, $\theta = \arctan(y/x)$, and $J_0(k\rho)$ is the zeroth-order Bessel function of the first kind. The electric potential energy of this configuration can then be expressed as
\begin{equation}
    \label{Upot}
    U(z_0,\ltf) = \frac{q^2}{4\epsilon^{(\text{sol})}z_0} \left[\mathcal{I}_1 -\epsilon^{(\text{sol})}\mathcal{I}_2 -1\right]
\end{equation}
where ($\mathcal{I}_1, \mathcal{I}_2$) are integrals over a dimensionless, dummy integration variable $\tau$
\begin{equation}
    \mathcal{I}_1 = \int_0^\infty d\tau \frac{4\left(\frac{\epsilon^{(\text{sol})}}{\epsilon^{(\text{el})}}\right)\tau e^{-2\tau}}{\sqrt{\tau^2+\left(\frac{z_0}{\ltf}\right)^2}+\left(\frac{\epsilon^{(\text{sol})}}{\epsilon^{(\text{el})}}\right)\tau}
\end{equation}
and
\begin{equation}
    \mathcal{I}_2 = \int_0^\infty d\tau \frac{2\left(\frac{z_0}{\ltf}\right)^2\tau e^{-2\tau}}{\left(\sqrt{\tau^2+\left(\frac{z_0}{\ltf}\right)^2}+\frac{\epsilon^{(\text{sol})}}{\epsilon^{(\text{el})}}\tau\right)^2\sqrt{\tau^2+\left(\frac{z_0}{\ltf}\right)^2}}
\end{equation}
For our purposes, the complexity of the expressions above may be abstracted by making an analogy to the method of images, and re-writing Eq. \ref{Upot} as the effective interaction of $q$ with a fictitious image (point) charge in the electrode, located at $-z_0$:      
\begin{equation}
    \label{Uim}
    U(z_0,\ltf) = \frac{q^2\xi_{\ltf}(z_0,\epsilon^{(\text{sol})})}{4\epsilon^{(\text{sol})}z_0} 
\end{equation}
where the image charge scaling function, $\xi_{\ltf}(z_0,\epsilon^{(\text{sol})})$ has been defined as
\begin{equation}
    \xi_{\ltf}(z_0,\epsilon^{(\text{sol})}) = \mathcal{I}_1 -\epsilon^{(\text{sol})}\mathcal{I}_2 -1
\end{equation}
This function informs on the scaling (with respect to $q$) of the fictitious image point charge as a function of position $z_0$, screening length in the electrode $\ltf$, and the dielectric constants of both media. It can be easily verified that for $\ltf=0$
\begin{equation}
    \label{xi_metal}
    \xi_{0}(z_0,\epsilon^{(\text{sol})}) = -1
\end{equation}
for any $z_0$ and $\epsilon^{(\text{sol})}$, in agreement with the result appropriate for a perfect metal. On the other hand, in the limit $\ltf\to\infty$
\begin{equation}
    \xi_{\infty}(z_0,\epsilon^{(\text{sol})}) = \frac{\epsilon^{(\text{sol})}-\epsilon^{(\text{el})}}{\epsilon^{(\text{sol})}+\epsilon^{(\text{el})}} 
\end{equation}
which is the result appropriate for the polarization that manifests at the dielectric discontinuity between two insulating materials. We will take $\epsilon^{(\text{el})} = 1$. This is the exact continuum result under the sharp boundary assumption. Since $\xi_{\ltf}(z_0,\epsilon^{(\text{sol})})$ is not a closed-form expression, its exact behavior for intermediate values of $\ltf$ is difficult to determine by simple inspection but it can be studied numerically, or through simulation by simply recording the electric potential energy between a test charge and the electrode at various values of screening length.

\begin{figure}[t]
    \centering
    \includegraphics[width=0.45\textwidth]{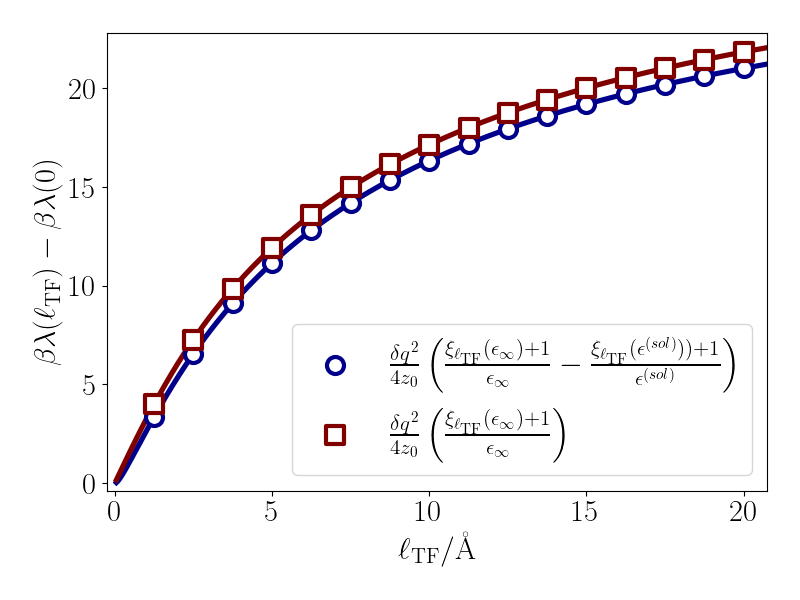}
    \caption{Dielectric continuum prediction of the reorganization energy as a function of screening length, with and without the term containing the static dielectric constant of the solvent}
    \label{fig:activation_figure}
\end{figure}

This basic insight has important implications for the rate of electron transfer. Consider now the solvent to be composed of an explicit collection of polar molecules. The exact expression for the electric potential energy as a function of the configuration of the system will depend on the configuration of all the electrolyte degrees of freedom, and it will be subject to thermal fluctuations.
In Marcus theory, within linear response the expectation value of the energy gap is completely determined by the reorganization energy and the driving force of the reaction
\begin{align}
        \label{reorg_connection}
        \langle\Delta E(z_0 , \ltf)\rangle_A &= -\langle\Delta E(z_0 , \ltf)\rangle_D\notag \\
        &= \lambda(z_0 , \ltf) + \Delta F(z_0 , \ltf)
\end{align}
where the average is over the fluctuations of the solvent. Under expectation values, we can approximate the fluctuations of the solvent with dielectric continuum theory. We  estimate the ensemble average of the gap by considering that the interaction between the instantaneously formed excess charge $\delta q$ and its image in the electrode will only be shielded by the fast (electronic) degrees of freedom of the solvent, characterized by the optical dielectric constant $\epsilon_{\infty}^{(\text{sol})}$. Under such approximations,  $\langle\Delta E(z_0 , \ltf)\rangle_A$ can be expressed as
\begin{equation}
    \begin{split}
        &\langle\Delta E(z_0 , \ltf)\rangle_A = \frac{\left(q_D^2-q_A^2\right) \xi_{\ltf}(z_0,\epsilon^{(\text{sol})})}{4\epsilon^{(\text{sol})}z_0} \\
        &+ \frac{\delta q^2}{4z_0}\left(\frac{\xi_{\ltf}(z_0,\epsilon^{\text{(sol)}}_{\infty})}{\epsilon^{\text{(sol)}}_{\infty}}-\frac{\xi_{\ltf}(z_0,\epsilon^{\text{(sol)}})}{\epsilon^{\text{(sol)}}}\right)
    \end{split}
\end{equation}
where first term is the reversible work of charging the redox species from $q_A$ to $q_D$. We can identify it as a contribution to the driving force of the reaction, $\Delta F$. The rest can be identified as the reorganization energy, $\lambda$. Note that the electrostatic contribution to the driving force is shielded by the solvent's static dielectric constant, possibly explaining why we observe such low values of driving force in a polar solvent such as water. The reorganization energy is 
\begin{equation}
    \label{lambda_ellTF}
    \begin{split}
        \lambda(z_0 , \ltf) =& \frac{\delta q^2 }{4z_0} \left(\frac{\xi_{\ltf}(z_0,\epsilon^{\text{(sol)}}_{\infty})}{\epsilon^{\text{(sol)}}_{\infty}}-\frac{\xi_{\ltf}(z_0,\epsilon^{\text{(sol)}})}{\epsilon^{\text{(sol)}}}\right) \\
        &+ \lambda_{\mathrm{B}}(z_0) 
    \end{split}
\end{equation}
the additional term $\lambda_{\mathrm{B}}(z_0)$ corresponds to a pure solvent contribution to the reorganization energy that is independent of screening in the electrode. It is associated to the difference in the self-energy of the redox species between a vertically excited state and equilibrium. Although independent of the electrode's metallicity, it can in principle be influenced by the existence of a boundary because the modes of the dielectric polarization field that couple to the charge would be modified in proximity to the interface, so it can still depend on the position of the redox species relative to the interface. However, in agreement with the `sharp-boundary' approximation, this geometric effect is expected to be small in comparison to pure electrostatic contributions, so it is reasonable to compute $\lambda_{\mathrm{B}}(z_0)$ with Marcus' dielectric continuum estimate of the reorganization energy for bulk reactions
\begin{equation}
    \label{Marcus_homo_diel_reorg}
    \lambda_{\mathrm{B}} \sim \frac{\delta q^2}{2a}\left(\frac{1}{\epsilon^{\text{(sol)}}_{\infty}}-\frac{1}{\epsilon^{\text{(sol)}}}\right)
\end{equation}
where $a$ is the radius of the redox ion. Recalling Eq. \ref{xi_metal}
\begin{equation}
    \lambda(z_0 , 0) = -\frac{\delta q^2}{4z_0}\left(\frac{1}{\epsilon^{\text{(sol)}}_{\infty}}-\frac{1}{\epsilon^{\text{(sol)}}}\right) + \lambda_{\mathrm{B}}(z_0)
\end{equation}
which reduces to the well-known dielectric continuum estimate from Marcus if we substitute the expression for $\lambda_{\mathrm{B}}(z_0)$ presented in Eq. \ref{Marcus_homo_diel_reorg}
\begin{equation}
    \label{reorg_perfect_metal}
    \lambda(z_0 , 0) = -\frac{\delta q^2}{2}\left(\frac{1}{\epsilon^{\text{(sol)}}_{\infty}}-\frac{1}{\epsilon^{\text{(sol)}}}\right)\left(\frac{1}{a}-\frac{1}{2z_0}\right) \, .
\end{equation}
In a polar solvent, the first term of Eq. \ref{lambda_ellTF} will be much larger than the second. Therefore, it would be reasonable to suggest approximating
\begin{equation}
    \lambda(z_0 , \ltf) - \lambda(z_0 , 0) \approx \frac{\delta q^2 }{4z_0} \frac{\left(\xi_{\ltf}(z_0,\epsilon^{\text{(sol)}}_{\infty})+1\right)}{\epsilon^{\text{(sol)}}_{\infty}}
\end{equation}
where Fig. \ref{fig:activation_figure} shows the continuum prediction of $\lambda$ for an ion placed $5 \mathrm{\AA}$ away from the electrode including and neglecting this term, It is clear that the difference is indeed very small. Even for very large values of screening length, it never exceeds $1 \kB T$. \\

\section{Comment on Dimensionality}

Strictly speaking, the Thomas-Fermi equations that were just presented in the main text are  valid for three-dimensional bulk systems. It is well known that electrostatic screening is generally weaker in 2D than in 3D \cite{noori_dielectric_2019}. It is therefore reasonable to question the choice of mapping the electrostatics of a 2D system to a three-dimensional theory. Firstly, we point out that a lot the same ideas translate with very little modification to the theory of screening in 2D \cite{ando_electronic_1982}. In the idealized case of a perfectly two-dimensional material confined to the x-y plane in contact with media that have dielectric constants $\epsilon_L$ for $z<0$ and $\epsilon_R$ for $z>0$, the screened Coulomb potential is given fully in terms of a Fourier-Bessel expansion of the form
\begin{equation}
    \label{2D-potential}
    \phi(\mathbf{r},z) = \int _0^\infty k \hat{\phi}(k,z) J_0(kr) dk
\end{equation}
in cylindrical coordinates. The Fourier-Bessel transform of the potential on the plane in response to a test charge placed at $\mathbf{r}=0$ and $z=z_0\geq0$ is given by
\begin{equation}
    \label{Fourier-Bessel-coeff}
    \hat{\phi}(k,0) = \frac{1}{\bar{\epsilon}}\frac{e^{-kz_0}}{k+\ell_{2D}^{-1}}
\end{equation}
where $\bar{\epsilon} = (\epsilon_L + \epsilon_R)/{2} $ and $\ell_{2D}$ is the effective screening length, defined as
\begin{equation}
    \label{2Dlength}
    \ell_{2D} = \frac{\bar{\epsilon}}{2\pi e^2 \left(\partial\rho/\partial \mu\right)} \approx \frac{\bar{\epsilon}}{2\pi e^2 D(\mu)}
\end{equation}
We can see that the screening length in 2D and 3D only differ by an overall square root. This means that Thomas-Fermi theory is still well defined in two-dimensions, and the connection between density of states and screening length still exists and is very similar to 3D. Conclusions drawn for a system modeled with the three-dimensional theory can therefore be used to make meaningful statements about screening in systems of lower dimension.  \\


\section{Model Density of States in TBG}


\begin{figure}[t]
            \includegraphics[width=8.5cm]{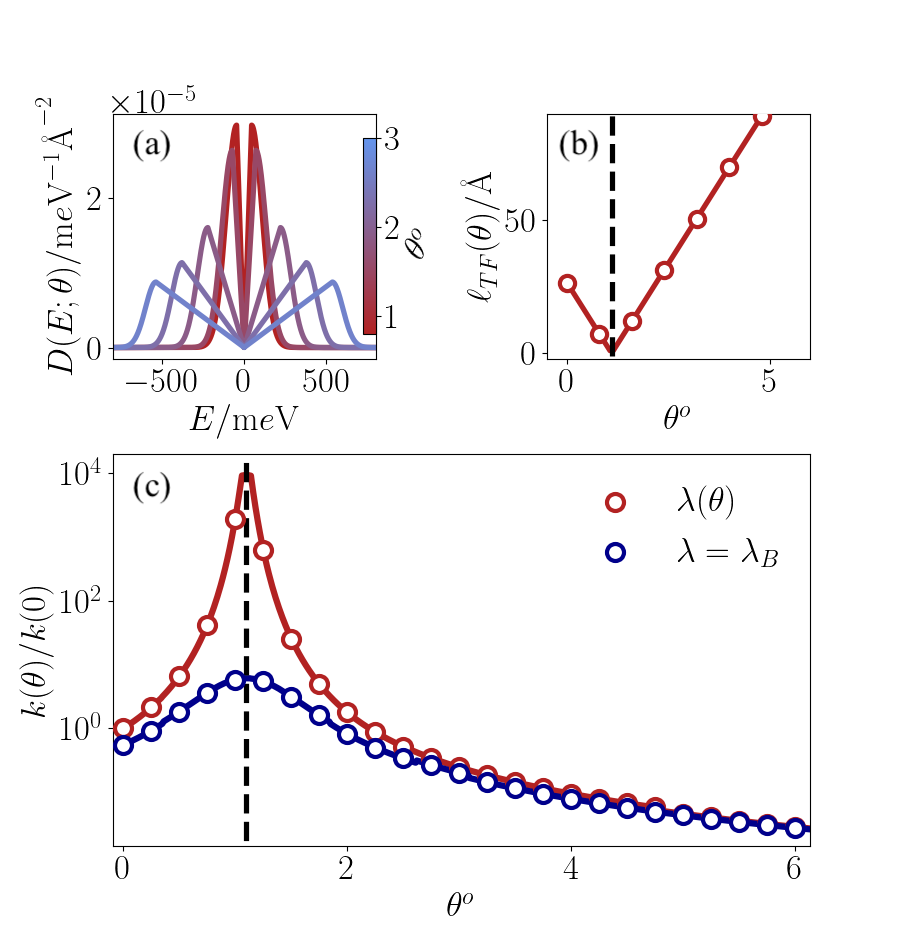}
    \caption{(a) Density of states for various twist angles for an empirical linear model (b) Screening length computed by evaluating the density of states in (a) at a finite Fermi energy ($4\times 10^{-5} \mathrm{E_h} \approx 1 \mathrm{meV}$) (c) Rate as function of twist angle evaluated using $D(E)$ and $\ltf$ in (a) and (b) (red line) and that with fixed $\lambda$ (blue line). The parameter $s = 9\times10^4 \ \mathrm{E_h}^{-2}$ was employed.}
    \label{fig:rate_angle}
\end{figure}

Here, we elaborate on the details and present alternative formulations of the model density of states in TBG used to obtain estimates of the angle-dependent rate.
We first discuss a linear model that is based on the fact that the band structure of a twisted graphene bilayer can be approximated as that of monolayer graphene with a renormalized Fermi velocity that vanishes at the magic angle. Close to the Dirac points, the density of states of a single graphene sheet is
\begin{equation}
    D(E) = \frac{2}{\pi \hbar^2 v_F^2} |E|
\end{equation}
where the Fermi velocity, $v_F$, of graphene is defined as the group velocity 
\begin{equation}
    \label{Fermi_vel}
    v_F = \nabla_k E(k) |_{k=K}
\end{equation}
and $K$ the Dirac point. 
Our proposal consists in making the Fermi-velocity angle-dependent, with the functional form of $v_F(\theta)$ having the basic requirement of vanishing at the first magic angle $\theta_m\approx 1.1^\text{o}$. We further assume that the rotation axis is centered at one of the carbon atoms, and $\theta=0$ corresponds to an AB Bernal stacked bilayer, implying that any angle-dependent property will at least be $2\pi/3$ symmetric. This leads to
\begin{equation}
    v_F(\theta) = v_0\left|\sin(\frac{3(\theta-\theta_m)}{2})\right|
\end{equation}
Where $v_0 \approx 0.45 \  \mathrm{a.u}$ is the Fermi velocity in monolayer graphene. To make the density of states integrable, we also introduce rapid decay after a cutoff on either side. The value of the cutoff is determined by the normalization requirement
\begin{equation}
    \label{DOSnormalization}
    \int_{-\infty}^{\infty} D(E;\theta) dE = \frac{N}{L^2}
\end{equation}
where $N$ is the number of electrons per Moiré unit cell and $L$ is the lattice constant. In principle, the size of the Moiré unit cell is a function of twist angle. 
This suggests that we should consider an angle-dependent lattice constant. Experimentally it has been observed that lattice relaxation effectively pins the lattice constant to a constant value for small angles.Therefore, we will consider the Moiré unit cell size to be approximately constant, and set it to its value at $\theta_m=1.1^\circ$. Since we are aiming to model the active bands of TBG, we set $N=8$, accounting for spin and valley degeneracy in both layers. The full model DOS then has the form
\begin{equation}
    \label{linearDOS}
    D(E;\theta) = 
    \begin{cases}
    \frac{1}{\hbar^2v_F^2(\theta)}|E| \qquad \qquad |E| \leq E_c \\
    \frac{E_c}{\hbar^2v_F^2(\theta)}e^{-s\left(E-E_c\right)^2} \qquad  \qquad E > E_c\\
    \frac{E_c}{\hbar^2v_F^2(\theta)}e^{-s\left(E+E_c\right)^2} \qquad  \qquad E < -E_c
    \end{cases}
\end{equation}
with $s>0$ setting the scale of the exponential decay, and the cutoff determined from normalization as
\begin{equation}
    E_c(\theta) = \frac{1}{2}\left(\sqrt{\frac{\pi}{s}+4\frac{N\hbar^2v_F^2(\theta)}{L^2}}-\sqrt{\frac{\pi}{s}}\right) 
\end{equation}
A plot of this model DOS is shown in Fig. \ref{fig:rate_angle}(a). Given that graphene is a semi-metal, this model is also conferring a semi-metallic character to TBG, evidenced by the fact that the density of states evaluated at the charge neutrality point ($E=0$) is identically zero for all values of screening length. This raises a concern because a vanishing density of states implies that the screening length would be infinity for all values of twist angle. Note, however, that the Fermi level will generically be shifted away from 0 by electrostatic gating, and if the DOS is instead evaluated an arbitrarily small energy away from the charge neutrality point, one recovers a well defined dependence of the screening length on twist angle. 

The TBG sheet in contact with an aqueous electrolyte on one side and with a complex layered material on the other side will be effectively gated, which justifies evaluating the density of states away from $E=0$. One way of addressing this is by invoking the concept of quantum capacitance, defined as the change in charge of the material with respect to changes in the chemical potential. The value of quantum capacitance for an untwisted double layer graphene system at zero applied potential has been estimated to be in the order of $2.5 \mu F \text{cm}^{-2}$ for typical values of electron density \cite{parhizgar2017quantum}. Nonzero quantum capacitance implies that the chemical potential is shifted away from charge neutrality, and in our model it translates to a chemical potential of roughly $0.042 \kB T$. 


A plot of the screening length as a function of twist angle from the equation above is shown in Fig. \ref{fig:rate_angle}(b). The behavior of the rate of interfacial electron transfer using this model is shown in Fig. \ref{fig:rate_angle} (c). Shown in the figure is the rate obtained from an angle-dependent reorganization energy (red curve), compared against the rate that results from considering an angle-independent reorganization energy (i.e. $\lambda=\lambda_{\mathrm{B}}$, dark blue curve). \\
The rate calculations were carried out by integrating the Marcus rate expression
\begin{equation}
    \label{rate_angle}
    \begin{split}
        k(\theta) =  \frac{2\pi}{\hbar} |V_{DA}|^2 \int_{-\infty}^\infty & D(E;\theta) f(E) \\
        &\times \frac{e^{-\beta\frac{\left(E-\left(\lambda(\ell_{TF}(\theta))+\Delta F\right)\right)^2}{4\lambda(\ell_{TF}(\theta))}}}{\sqrt{4\pi k_BT\lambda(\ell_{TF}(\theta))}} dE 
    \end{split}
\end{equation}
where the function $\lambda(\ltf(\theta))$ was obtained by fitting the behavior of the image charge scaling function to an expression of the form $\xi_{\ell_{TF}} \sim a/(b+c \ell_{TF}^2)$. Our findings indicate that, to a good approximation, the dependence of the reorganization energy on screening applies universally for all redox species in a given solvent that exchange the same number of electrons. Therefore, the reorganization energy only depends on the chemical identity of the redox system through the bulk term $\lambda_\mathrm{B}$. In order to establish a closer connection to the experimental report, throughout this work we used the value of $\lambda_\mathrm{B}$ appropriate for the \ch{[Ru(NH_3)_6]^{2+}/[Ru(NH_3)_6]^{3+}} redox couple, of 0.82 eV (32.8 $\kB T$) \cite{yu_tunable_2022}. 

The double Lorentzian model presented in the main text is motivated by experimental measurements and numerical results that suggest that bilayer graphene systems actually have a finite DOS at the charge neutrality point \cite{wong_cascade_2020, yu_tunable_2022,rozhkov_electronic_nodate, lopes_graphene_2007}. The explicit functional form of this model is
\begin{eqnarray}
    \label{LorentzianModel}
        D(E;\theta) &= \frac{N \gamma}{4\pi L^2(\theta)}\left[\frac{1}{\left(E-a|\sin(\frac{(\theta-\theta_m)}{T(\theta)})|\right)^2+(\gamma/2)^2}+ \right . \nonumber \\ &\left . \frac{1}{\left(E+a|\sin(\frac{(\theta-\theta_m)}{T(\theta)})|\right)^2+(\gamma/2)^2}\right]  
\end{eqnarray}
constructed to satisfy the normalization in Eq. \ref{DOSnormalization}, where $T(\theta)=c+b\theta$. Since this DOS evaluates to nonzero values at $E=0$, we can employ the usual definition of the screening length, 

\begin{equation}
    \ell_{TF}(\theta) = \sqrt{\frac{L_z L^2}{2 N \gamma} \left(\left(a|\sin((\theta-\theta_m)/T(\theta))|\right)^2+(\gamma/2)^2\right)}
\end{equation}

where $L_z$ is a length scale in the $z$-direction, which we will set to 1. As shown in the main text, the overall behavior of the rate under this model is quite similar to the linear model, with the exceptions that (1) the maximal rate enhancement with respect to $\theta=0$ is now only of roughly one order of magnitude, and (2) the behavior close to the magic angle is smoother than in the previous model. Both of these differences, along with the non-vanishing DOS at the Fermi level, make this model more consistent with experimental data. The results of the angle-dependent rate shown in the main text for this model correspond to a choice of parameters of $a = 0.12 \ \mathrm{E_h}$, $\gamma = 2\times 10^{-3} \ \mathrm{E_h}, c=1, b=20$. It should be noted, however, that although we present results for a specific choice of parameters, the superior rate enhancement that results from considering an angle-dependent reorganization energy is a robust feature that manifests across a wide range of choices. 

\begin{figure}[t]
    \centering
    \includegraphics[width=0.45\textwidth]{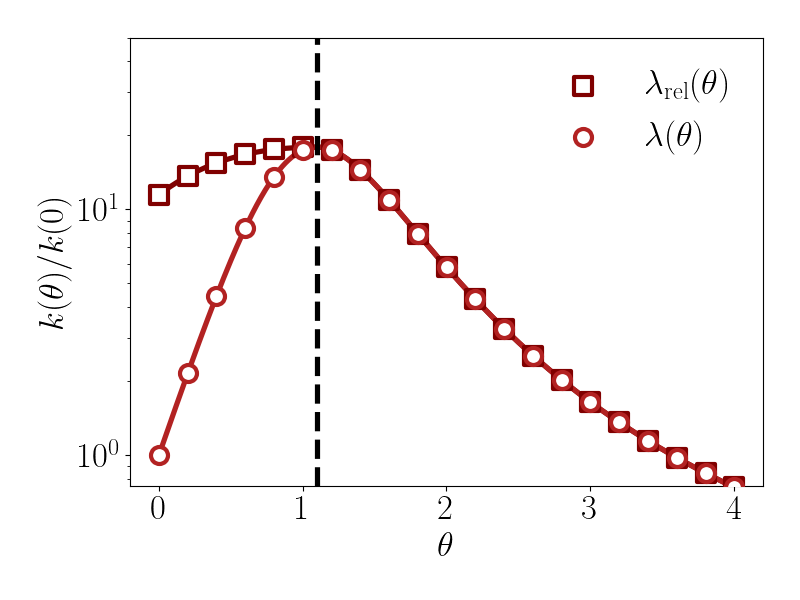}
    \caption{Rate as a function of twist angle with a modified reorganization energy $\lambda_{\mathrm{rel}}(\theta)$ to approximately account for lattice relaxation effects at small angles.}
    \label{fig:lattice_relax}
\end{figure}

Experimentally, the rate  for $\theta<\theta_m$ only seems to decrease slowly away from the magic angle. This effect is understood to be a consequence of lattice relaxation, whereby local strain in the AA stacking domains `fixes' the local twist angle to be of roughly $1.2^o$ for any global twist angle of less than $1.2^o$. Therefore, for small angles, the local electrochemical activity remains constant, and changes in the global rate are only caused by modifications in the size and relative concentration of AA vs AB/BA stacking domains, which are determined by the interplay of global twist and superlattice reconstruction. This feature does not emerge in our description, as we do not account for lattice relaxation in any way. A crude way of incorporating this effect is to consider that the reorganization energy is not affected by the area fraction of stacking domains, but rather is informed only by the local value of twist angle in the vicinity of the redox species. This local electrochemical activity is seen to remain constant for small angles, so we propose setting $\lambda(\theta)$ to the constant $\lambda(\theta_m)$ for $\theta<\theta_m$, and then using its usual angle dependence for $\theta>\theta_m$. The global density of states, on the other hand, implicitly accounts for the aforementioned changes in area fraction of stacking domains, so it is kept unchanged. The result of applying this scheme for the double-Lorentzian model DOS is shown in Fig. \ref{fig:lattice_relax}, where indeed we can see that we are able to capture a more subtle angle dependence at small angles, and a steeper descent in electrochemical activity at larger angles. Although this seems like a somewhat \textit{ad hoc} modification, we believe that it is supported by physically sound assumptions. In particular, since we have seen that the reorganization energy mainly depends on twist angle through modified charge polarization effects in the electrode, and this polarization only manifests locally in the vicinity of the redox species, it makes sense that if the \textit{local} electrochemical activity remains constant due to lattice relaxation effects, so will the reorganization energy of the reaction. \\

\section{Rates of energetically misaligned redox couples}

\begin{figure}[b]
    \centering
    \includegraphics[width=0.45\textwidth]{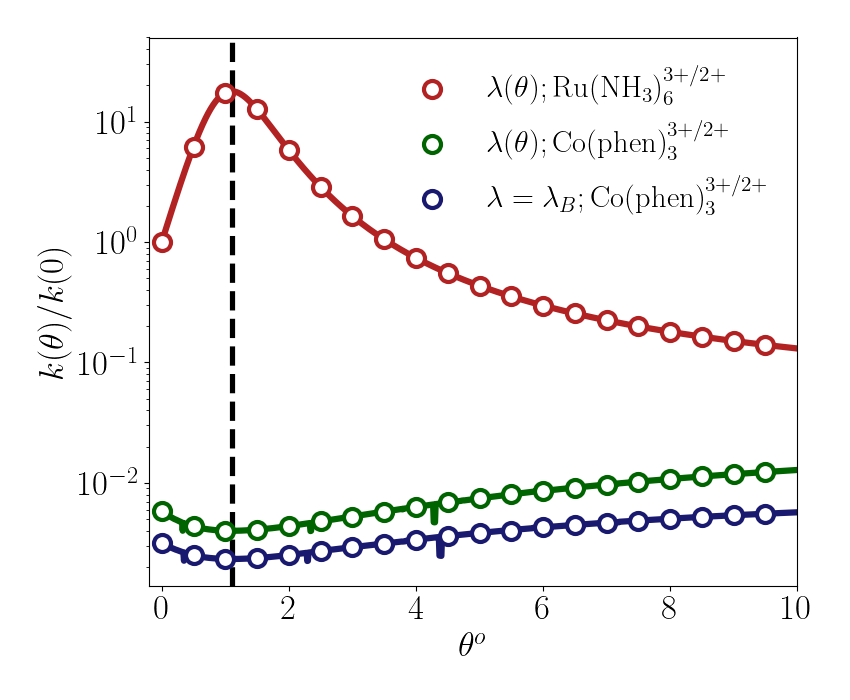}
    \caption{Angle-dependent rates for a reaction with a standard redox potential misaligned with the flat bands of TBG. Achieved by shifting the Fermi level by ~0.3 eV }
    \label{fig:rate_cobalt}
\end{figure}

In the original study \cite{yu_tunable_2022}, it was found that if the redox couple \ch{Co(phen)_3^{2+/3+}} was used instead of \ch{Ru(NH_3)_6^{2+/3+}}, much lower rates were observed, and the rate enhancement as a function of twist angle was significantly reduced. As surmised in the original work, this is likely due to a misalignment of the redox potential of this species with the flat-band region of the electronic states in TBG. This misalignment can be approximately captured by shifting the Fermi level and the energy gap distributions with respect to the flat-band region of the density of states, and correspondingly modifying the screening length. In Fig.\ref{fig:rate_cobalt}, we illustrate the rates that result of this procedure. The overall rate is 1-2 orders of magnitude lower, the angle dependence is much weaker, and considering an angle-dependent reorganization energy increases the rate, but not significantly.  

\section{Rates with Tight-Binding Density of States}

In this work, we employed empirical models to describe the angle-dependent structure of the density of states in TBG. An alternative to this approach is to employ the density of states derived from tight-binding calculations under the Bistritzer-MacDonald model\cite{bistritzer_moire_2011} for the band structure of TBG, as implemented previously\cite{carr2019exact, carr2020electronic, fang2016electronic}. Here we show the comparison of the rates obtained with our model DOS, and those computed from the DOS derived from tight-binding calculations.\\

\begin{figure}[t]
    \centering
    \includegraphics[width=0.4\textwidth]{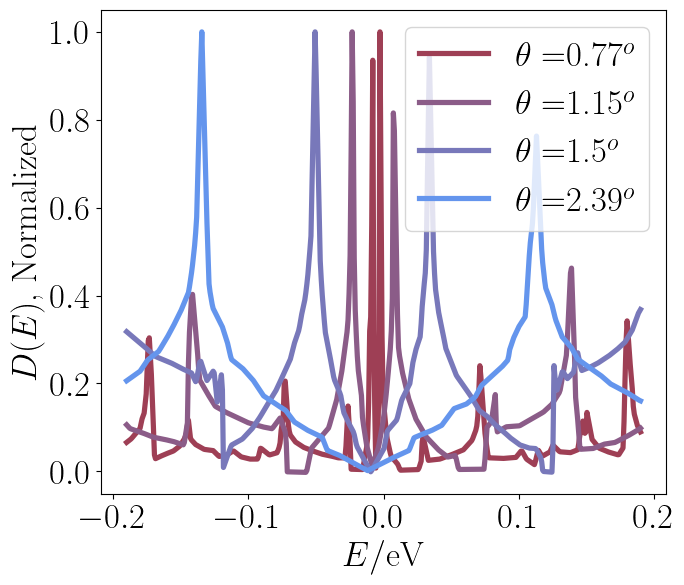}
    \caption{Density of states derived from tight-binding calculations for various twist angles. Shown are the DOS normalized by their maximum value.}
    \label{fig:rate_tight_binding}
\end{figure}

In order to integrate the rate expression, a function of the tight-binding DOS was constructed from the calculation data through simple linear interpolation, and the integral was performed only over the domain $[-0.2,0.2]$ eV, verifying that it had numerically converged. There is a large body of experimental evidence suggesting that the DOS at the Fermi level in bilayer graphene systems is actually finite, but the tight-binding calculations exhibit a node at E=0. To address this, the screening length was derived from the tight-binding results in exactly the same way as was done for the linear model presented previously, by evaluating the DOS a small energy away from charge neutrality ($\sim$ 1 meV). Figure ~\ref{fig:rate_tight_binding} compares the rates that result from this procedure with those obtained using the double-Lorentzian model, and the experimental rates. They all agree reasonably well, and the agreement with experiment crucially depends on including the angle-dependent correction to the reorganization energy.

\begin{figure}[t]
    \centering
    \includegraphics[width=0.45\textwidth]{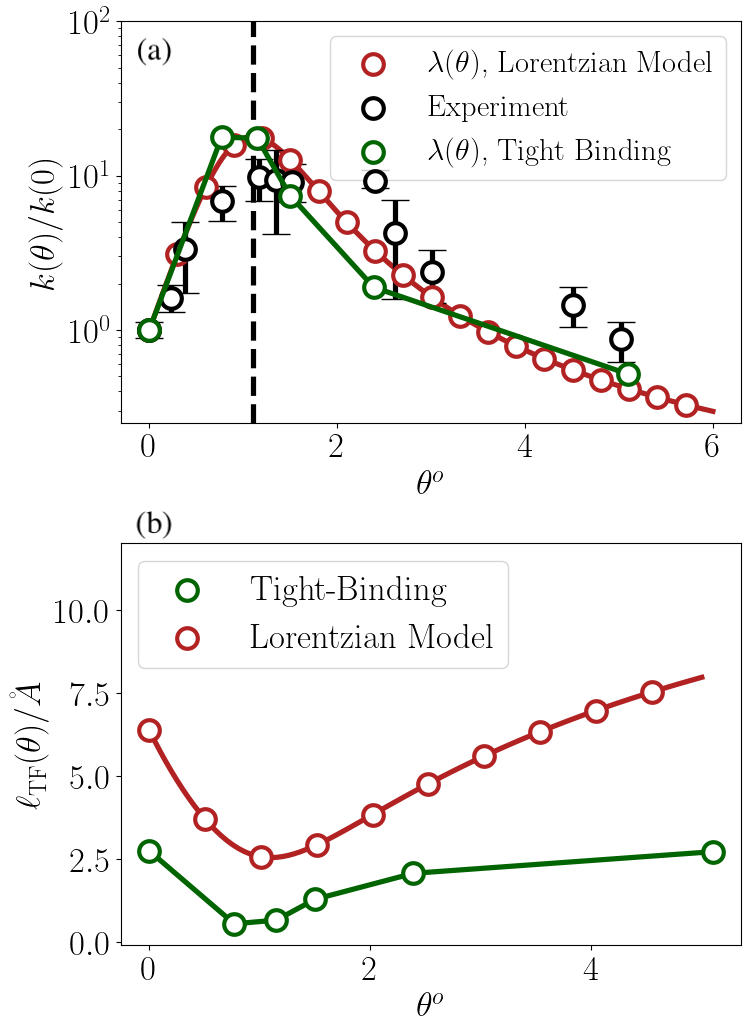}
    \caption{(a) Comparison of the experimental interfacial electron transfer rate with those calculated using DOS derived from tight-binding caluclations and the empiricial model presented in this work. All calculations account for angle-dependent reorganization energy through screening-modulated image interactions (b) TF screening length as a function of twist angle for the tight-binding and emprirical DOS}
    \label{fig:rate_tight_binding}
\end{figure}

\section{References}

\bibliography{references,ref2}